\documentclass[10pt, prd, twocolumn, nofootinbib,preprint,superscriptaddress]{revtex4}
\pdfoutput=1
\usepackage[T1]{fontenc}
\usepackage{amsmath,amssymb}
\usepackage{epsfig}
\usepackage{graphicx}
\usepackage[usenames,dvipsnames]{color}
\usepackage{subfigure}
\usepackage{slashed}
\usepackage[colorlinks,citecolor=blue]{hyperref}
\usepackage{pdfpages}
\usepackage{color}
\usepackage{epstopdf}
\usepackage{graphicx}
\usepackage{dcolumn}
\usepackage{bm}
\usepackage{hyperref}
\usepackage{braket}

\begin{document}


\title{Effects of gravitational lensing on neutrino oscillation in $ \gamma $-spacetime}

\author{Hrishikesh~Chakrabarty}
\email{chrishikesh17@fudan.edu.cn}
\affiliation{School of Astronomy and Space Sciences, University of Chinese Academy of Sciences, Beijing, China}

\author{Debasish Borah}
\email{dborah@iitg.ac.in}
\affiliation{Department of Physics, Indian Institute of Technology Guwahati, Assam 781039, India}

\author{Ahmadjon Abdujabbarov}
\email{ahmadjon@astrin.uz}
\affiliation{Shanghai Astronomical Observatory, 80 Nandan Road, Shanghai 200030, P. R. China}
\affiliation{Ulugh Beg Astronomical Institute, Astronomy St 33, Tashkent 100052, Uzbekistan}
\affiliation{National University of Uzbekistan, Tashkent 100174, Uzbekistan}
\affiliation{Tashkent Institute of Irrigation and Agricultural Mechanization Engineers,39 Kori Niyoziy st., Tashkent 100000, Uzbekistan}

\author{Daniele~Malafarina}
\email{daniele.malafarina@nu.edu.kz}
\affiliation{Department of Physics, Nazarbayev University, 53 Kabanbay Batyr avenue, 010000 Astana, Kazakhstan}


\author{Bobomurat~Ahmedov}
\email{ahmedov@astrin.uz}
\affiliation{Ulugh Beg Astronomical Institute, Astronomy St 33, Tashkent 100052, Uzbekistan}
\affiliation{National University of Uzbekistan, Tashkent 100174, Uzbekistan}
\affiliation{Tashkent Institute of Irrigation and Agricultural Mechanization Engineers,39 Kori Niyoziy st., Tashkent 100000, Uzbekistan}

\date{\today}

\begin{abstract}
We study the effects of gravitational lensing on neutrino oscillations in the $\gamma$-spacetime which describes a static, axially-symmetric and asymptotically flat solution of the Einstein's field equations in vacuum. Using the quantum-mechanical treatment for relativistic neutrinos, we calculate the phase of neutrino oscillations in this spacetime by considering both radial and non-radial propagation. We show the dependence of the oscillation probability on the absolute neutrino masses, which in the two-flavour case also depends upon the sign of mass squared difference, in sharp contrast with the well-known results of vacuum oscillation in flat spacetime. We also show the effects of the deformation parameter $\gamma$ on neutrino oscillations and reproduce previously known results for the Schwarzschild metric. 
We then extend these to a more realistic three flavours neutrino scenario and study the effects of the parameter $\gamma$ and the lightest neutrino mass while using best fit values of neutrino oscillation parameters.

\end{abstract}

\maketitle


\section{\label{sec:intro}Introduction}
The fact that neutrinos have tiny masses and large mixing has been well-established by now, thanks to several experimental evidences gathered over a long periods of time \cite{Zyla:2020zbs}. For a latest update on neutrino parameter values, one may have a look at the latest global fit of neutrino data \cite{Esteban:2020cvm, deSalas:2020pgw}. In spite of this great success story, there are still several unanswered questions related to neutrinos. In addition to the fundamental question of the origin of light neutrino mass, we do not know a few things from experimental point of view. For example, we still do not know whether the atmospheric mixing angle lies in the first or second octant, even though the best fit value prefers the second octant. Also, since only two mass squared differences are experimentally measured, we do not have any idea about the lightest neutrino mass leading to the possibility of both normal ordering (NO): $m_1 < m_2 < m_3$ as well as inverted ordering (IO): $m_3 < m_1 < m_2$. The global fit data, however, suggests a normal mass ordering. While neutrino oscillation experiments are sensitive to mass squared differences only, there are other ways to measure the absolute neutrino mass scale. For example, the tritium beta decay experiment KATRIN has recently measured an upper limit of 1.1 eV on absolute neutrino mass scale \cite{KATRIN:2019yun} which is however, weaker than the upper limits from cosmology.
Latest data from Planck collaboration constrains the sum of absolute neutrino masses as $\sum_i \lvert m_i \rvert < 0.12$ eV \cite{Aghanim:2018eyx}. In spite of recent advances in determining the leptonic Dirac CP phase \cite{Abe:2019vii} suggesting a maximal CP violation, the measurement of leptonic Dirac CP phase is not statistically significant enough to be considered as a discovery.

The results of various neutrino oscillation experiments are interpreted by using the quantum mechanical neutrino oscillation probability derived in flat Minkowski spacetime. While such descriptions are valid for the laboratory based oscillation experiments, in extreme astrophysical environments or cosmology in general, the effects of non-trivial spacetime background may be important. This has led to several theoretical studies on the effects of curved spacetime on neutrino oscillations \cite{Wudka:1991tg,Grossman:1996eh, Cardall:1996cd, Piriz:1996mu, Fornengo:1996ef, Ahluwalia:1996ev,Bhattacharya:1999na, Pereira:2000kq,Crocker:2003cw, Lambiase:2005gt, Godunov:2009ce, Ren:2010yf, Geralico:2012zt, Chakraborty:2013ywa, Visinelli:2014xsa,Chakraborty:2015vla,Zhang:2016deq, Alexandre:2018crg, Blasone:2019jtj, Buoninfante:2019der, Boshkayev:2020igc, Mandal:2021dxk, Koutsoumbas:2019fkn, Swami:2020qdi, Capolupo:2020wlx}. The general conclusion reached in most of these works is the increase in effective neutrino oscillation length due to gravitational redshift in such curved spacetime. In this work, we consider the effects of gravitational lensing on neutrino oscillations. In the presence of lensing around a massive astrophysical object, the oscillation probability at a particular point is calculated by considering all possible trajectories of neutrinos which get focused at that point due to the lensing phenomenon. While this was first discussed in \cite{Fornengo:1996ef}, some follow-up works have also appeared in the literature \cite{Crocker:2003cw, Alexandre:2018crg, Swami:2020qdi}. While these works have focused on the Schwarzschild spacetime for such studies, we consider here a different and well-motivated spacetime metric. 

In this paper, we study gravitational lensing of neutrinos in the so-called $ \gamma $-metric, also known as the `Zipoy-Voorhees' spacetime \cite{gamma1,Voorhees:1970ywo}. The $ \gamma $-metric is a static, axially-symmetric and asymptotically flat solution of Einstein's equations in vacuum and it belongs to the Weyl class of exact solutions. The metric is characterized by only two parameters: $ M $ related to the mass of the source and $ \gamma $ related to the static mass quadrupole moment of the source. The parameter $ \gamma $ describes deviations from the spherical symmetry in such a way that $ \gamma = 1 $ reproduces the Schwarzschild metric while $\gamma>1$ ($\gamma<1$) describes an oblate (prolate) source. The structure of the metric at small radii is also very interesting since when $ \gamma \neq 0 $ the horizon at $ 2M $ is replaced by a true curvature singularity. Therefore, the singularity in the $ \gamma $-metric is naked and Virbhadra in \cite{Virbhadra:1996cz} showed that the singularity has `directional nakedness' which depends on the angle of approach \cite{Kodama:2003}. For $ \gamma < 1 $, this singularity is globally visible along polar as well as equatorial direction. For $ \gamma > 1 $, the singularity is globally naked along equatorial direction, but it is not visible (even locally) along the polar direction. In \cite{Chakrabarty:2017ysw}, it was shown that the same singularity at $ 2M $ can be resolved in conformal gravity theories. Also it is worth noticing that, the `Zipoy-Voorhees' metric is non-integrable, as shown in \cite{Lukes-Gerakopoulos:2012qpc}, which  leads to interesting and possibly chaotic behavior for the general motion of test particles.

Goedesic motion of massive and massless particles in the $ \gamma $-metric was studied in \cite{Herrera:1998, Chowdhury:2012, Boshkayev:2016, Toshmatov:2019bda}. Shadows and optical properties was calculated in \cite{Abdikamalov:2019ztb} and it was shown that this spacetime can behave as a black hole mimicker for some allowed values of $ \gamma $. Therefore, the $\gamma $-metric provides an excellent and simple candidate to study toy models of astrophysical scenarios where the exterior region of a massive compact object is not given by a black hole line element.
We calculate the quantum mechanical phase of neutrino oscillations in this metric by considering both radial and non-radial propagation of neutrinos. Using the weak-field approximation, we also arrive at simple expressions for this phase and use it to calculate the neutrino oscillation probabilities. We then use the effects of gravitational lensing and calculate the oscillation probabilities for both two-flavor and three-flavor regimes.

The article is structured as follows: In Sec.~\ref{sec:nuosc}, we give a very brief review of neutrino oscillation in flat and curved spacetimes and in Sec.~\ref{sec:gamma}, we present a brief overview of $ \gamma $-metric. 
In Sec.~\ref{sec:phase}, we calculate the phase for neutrino oscillations in the $ \gamma $-metric for radial and non-radial propagation of neutrinos and in Sec.~\ref{sec:oscprob}, we calculate the oscillation probability of neutrinos moving in a geometry described by $ \gamma $-metric. In Sec.~\ref{sec:lensing}, we discuss gravitational lensing of neutrinos and present the numerical results of both two and three flavor neutrino oscillation probability in this spacetime. 
Finally, in Sec.~\ref{sec:conc} we summarize and comment on our results. Throughout the paper we employ $ (-,+,+,+) $ signature for the line element and adopt geometrical units setting $ c = G = \hbar = 1 $.

\section{\label{sec:nuosc}Neutrino oscillations}

Neutrinos are produced and detected in different flavor eigenstates denoted by $ \ket{\nu_\alpha} $, and the flavor eigenstates are superposition of mass eigenstates denoted by $ \ket{\nu_i} $. So a flavor eigenstate can be written in terms of mass eigenstates as 
\begin{equation}
    \ket{\nu_\alpha} = \sum U^*_{\alpha i}\ket{\nu_i},
\end{equation}
where $ \alpha = e, \mu, \tau $ and $ i = 1, 2, 3 $. $ U $ is a $ 3\times3 $ unitary mixing matrix. For three flavor neutrino oscillation, this is known as the Pontecorvo-Maki-Nakagawa-Sakata (PMNS) leptonic mixing matrix \cite{Pontecorvo:1957qd, Maki:1962mu, Pontecorvo:1967fh}.

We assume that the neutrino wave-function is a plane wave as considered originally in \cite{Pontecorvo:1957qd, Maki:1962mu, Pontecorvo:1967fh} and it propagates from a source S located at a spacetime event $ (t_S, x_S) $ to a detector D located at a spacetime event $ (t_D,x_D) $. So the wave-function at the detector point is given by
\begin{equation}
    \ket{\nu_i(t_D,x_D)} = \exp\left( -i\Phi_i \right)\ket{\nu_i(t_S,x_S)}\ .
\end{equation}

Neutrinos are expected to be produced initially in the flavour eigenstate $ \ket{\nu_\alpha} $ at S and then travel to the detector D. In that case, the probability of the change in neutrino flavour from $ \nu_\alpha $ to $ \nu_\beta $ at D is given by
\begin{equation}
    \begin{aligned}
        \mathcal{P}_{\alpha\beta} &= |\braket{\nu_\beta | \nu_\alpha(t_D,x_D)}|^2 =\\
        &= \sum_{i,j}U_{\beta i}U_{\beta j}^*U_{\alpha j}U_{\beta i}^* \exp\left[ -i(\Phi_i - \Phi_j  ) \right].
    \end{aligned}
\end{equation}
In flat spacetime, the phase $ \Phi_i $ is given by
\begin{equation}\label{phaseflat}
    \Phi_i = E_i(t_D - t_S) - p_i(\vec{x}_D-\vec{x}_S). 
\end{equation}
It is typically assumed that all the mass eigenstates in a flavour eigenstate initially produced at the source have equal momentum or energy \cite{Bilenky:1978nj}.  Either of these assumptions together with $ (t_D-t_S) \simeq |(\vec{x}_D-\vec{x}_S)| $ for relativistic neutrinos ($ E_i \gg m_i $) lead to
\begin{equation}
    \Delta \Phi_{ij} \equiv \Phi_i - \Phi_j \simeq \frac{\Delta m_{ij}^2}{2E_0}(\vec{x}_D-\vec{x}_S),
\end{equation}
where $ \Delta m_{ij}^2 = m_i^2 - m_j^2 $, and $ E_0 $ is the average energy of the relativistic neutrinos produced at the source. 

To generalize the expression of the phase $ \Phi_i $ for neutrino propagation in curved spacetime, Eq.~\eqref{phaseflat} is written in the covariant form \cite{Fornengo:1996ef}
\begin{equation}
    \Phi_i = \int^D_S p_\mu^{(i)}dx^\mu,
\end{equation}
where 
\begin{equation}
    p_\mu^{(i)} = m_i g_{\mu\nu}\frac{dx^\mu}{ds}
\end{equation}
is the canonical conjugate momentum to the coordinates $ x^\mu $ and $ g_{\mu\nu} $ and $ ds $ are the metric tensor and line element of the curved spacetime, respectively. 

\section{\label{sec:gamma} The $ \gamma $-metric}

The $\gamma$-metric is an exact solution of the Einstein's field equations in vacuum belonging to the class of Weyl metrics. The line element for $\gamma$-metric can be written as \cite{gamma1,Voorhees:1970ywo}
\begin{equation}\label{e-gamma}
    \begin{aligned}
        ds^2 = &- f^\gamma dt^2 + f^{\gamma^2-\gamma}g^{1-\gamma^2} \left( \frac{dr^2}{f} + r^2 d\theta^2 
        \right)+ \\
        &+ f^{1-\gamma} r^2\sin^2\theta d\phi^2,
    \end{aligned}
\end{equation}
where
\begin{equation}
    \begin{aligned}
        & f = 1 - \frac{2M}{r}, \\
        & g = 1 - \frac{2M}{r} + \frac{M^2\sin^2\theta}{r^2}.
    \end{aligned}
\end{equation}

There are two parameters characterizing this line element: $M > 0$ is related to the mass of the source, and $\gamma > 0$ quantifies the departure of the geometry from spherical symmetry. For $\gamma = 1$, the spacetime becomes spherically symmetric and from Eq.~\ref{e-gamma} we recover the Schwarzschild solution in Schwarzschild coordinates. For $\gamma > 1$ ($\gamma < 1$), the spacetime posses oblate (prolate) structure, which may be used to describe the exterior of a compact object. Interiors for the $\gamma$-spacetime were obtained in \cite{Stewart:1982, Herrera:2005, Hernandez-Pastora:2016}. In order to have a better understanding of the interpretation of the parameters, it is useful to consider the asymptotic expansion in multipoles of the gravitational potential. We see that the total mass as measured by an observer at infinity $ M_{\rm tot} $, namely the monopole moment, and the quardupole moment $ Q $ are given by \cite{HPM}
\begin{equation}
    \begin{aligned}
         M_{\rm tot} &= M_{\rm ADM} = \gamma M \ , \\
         Q &= \frac{\gamma M^3(1-\gamma^2)}{3}.
    \end{aligned}
    \label{eq:ADM}
\end{equation}
For $\gamma \neq 1$, the spacetime has a naked singularity at the radial coordinate $ r_{\rm sing} = 2M $ as it can be seen by calculating the Kretschmann invariant. So the spacetime is geodesically incomplete with the radial coordinate range $ r \in (2M, \infty) $. In the following discussions, we consider the exterior of a massive object to be approximated by the $ \gamma $-metric in the mentioned radial coordinate range.

\section{\label{sec:phase}Phase of neutrino oscillation}
Let us now calculate the phase of neutrino oscillation probability in the $ \gamma $-spacetime. First, for simplicity, we shall rename the metric coefficients as follows.
\begin{equation}
    \begin{aligned}
         g_{tt} & = -A, \ \ \ \ g^{tt} = -1/A, \\
         g_{rr} & = B, \ \ \ \ g^{rr} = 1/B, \\
         g_{\theta\theta} & = C, \ \ \ \ g^{\theta\theta} = 1/C, \\
         g_{\phi\phi} & = D, \ \ \ \ g_{\phi\phi} = 1/D.
    \end{aligned}
\end{equation}
We can write the components of canonical momenta $  p_\mu^{(k)} $ for test particles moving on the equatorial plane $ \theta = \pi/2 $ in the $\gamma$-spacetime as
\begin{equation}\label{e-canmom}
    \begin{aligned}
        & p_t^{(k)} = m_k g_{tt}\frac{dt}{ds},  \\
        & p_r^{(k)} = m_k g_{rr}\frac{dr}{ds}, \\
        & p_\phi^{(k)} = m_k g_{\phi\phi}\frac{d\phi}{ds},
    \end{aligned}
\end{equation}\label{e-msrel}
where $m_k$ is the particle's mass, i.e. the neutrino mass eigenstate, ad $s$ is the trajectory's affine parameter. 
These momenta are related to each other and the mass of the $ k $-th eigenstate by the mass-shell relation
\begin{equation}
    -m_k^2 = g^{tt}p_t^2 + g^{rr}p_r^2 + g^{\phi\phi}p_\phi^2.
\end{equation}
To avoid confusion, we have dropped the prefix `$ k $' in the momenta.

\subsection{Radial case}

We first investigate the case of radial propagation of neutrinos. For  radial propagation, $ d\phi = 0 $. Therefore, from Eq.~\eqref{e-canmom}, we have
\begin{equation}\label{e-canmon2}
    \frac{dt}{ds} = \frac{p_t}{m_k g_{tt}}, \ \ \ \ \frac{dr}{ds} = \frac{p_r}{m_k g_{rr}}. 
\end{equation}
Thanks to the spacetime being static we know that the energy $E$ of test particles in conserved and we can define, $ p_t = -E_k $ and $ p_r = p_k(r) $, where the subscript $k$ refers to the neutrino flavor. With this notation, the phase of a neutrino propagating radially in a light-ray trajectory becomes 
\begin{equation}\label{eq-phase-basic}
    \Phi_k = \int^D_S \left[ -E_k\left( \frac{dt}{dr} \right)_0 + p_k(r) \right]dr,
\end{equation}
where, $ S $ and $ D $ denote the source of the neutrino and detector, respectively. We again emphasize that the phase in Eq.~\eqref{eq-phase-basic} is not the phase on a classical trajectory for the mass eigenstates but the phase calculated on a light-ray trajectory \cite{Fornengo:1996ef}. Now, from Eq.~\eqref{e-canmon2}, we get the light-ray differential as
\begin{equation}\label{e-lrdiff}
    \left(\frac{dt}{dr}\right)_0 = \frac{E_0}{p_0(r)} \frac{B}{A},
\end{equation}
where $ E_0 $ is the energy of a massless particle at infinity. The mass-shell relation gives
\begin{equation}\label{e-msrel2}
    \begin{aligned}
        & p_0(r) = \pm E_0\sqrt{\frac{B}{A}}, \\
        & p_k(r) = \pm \sqrt{\frac{BE_k^2}{A} - Bm_k^2}.
    \end{aligned}
\end{equation}
Using Eq.~(\ref{e-lrdiff}) and (\ref{e-msrel2}) in the expression for phase, we obtain
\begin{equation}
    \begin{aligned}
        \Phi_k 
        & = \pm \int^D_S E_k\sqrt{\frac{B}{A}} \left[ - 1 + \sqrt{ 1 - \frac{m_k^2A}{E_k^2} } \right]dr.
    \end{aligned}
\end{equation}
Now, using the fact that $ 0< A< 1 $, we expand the square-root under the bracket to write
\begin{equation}
    \Phi_k = \pm \int^D_S \sqrt{AB} E_k \frac{m_k^2}{2E_k^2} \ dr. 
\end{equation}
In relativistic approximation ($ m_k << E_k $), the following relation holds \cite{Fornengo:1996ef}
\begin{equation} \label{rel-approx}
    \begin{aligned}
        E_k & \simeq E_0 + \mathcal{O}\left(\frac{m_k^2}{2E_0}\right)\ , \\ 
        E_k \frac{m_k^2}{2E_k^2} & \simeq E_0 \frac{m_k^2}{2E_0^2}.
    \end{aligned}
\end{equation}
Hence the phase becomes
\begin{equation}
    \Phi_k = \pm \frac{m_k^2}{2E_0} \int^D_S \sqrt{AB} \ dr.
\end{equation}
Integrating the above integral from $ r_S $ to $ r_D $, we get
\begin{equation}\label{e-radphase}
    \Phi_k \simeq \pm \frac{m_k^2}{2E_0}(r_D-r_S)\left[ 1 - \frac{M^2(\gamma^2-1)}{2r_Sr_D} \right].
\end{equation}
Here the phase is calculated up to order $ \mathcal{O}\left(M^2/r^2\right) $ to obtain some information on its dependence on $\gamma$. In fact, we can see that $ \gamma $ has no effect in the order $ \mathcal{O}\left( M/r \right) $. So in the lowest order in ($ M/r $), the phase is same as that in Schwarzschild spacetime. This behaviour is interesting as the phase of the radially propagating neutrinos in the equatorial plane and in the weak field limit is not affected by the deformed geometry of the spacetime. The reason behind this is that, in the weak field limit $ g_{tt}g_{rr} \simeq 1 $ for the $ \gamma $-metric.

\subsection{Non-radial case}
Now we would like to concentrate on non-radial motion of neutrinos. Let us write the phase for a neutrino travelling along a light ray using Eq.~\eqref{e-canmom} and Eq.~\eqref{e-msrel2} as
\begin{equation} 
    \Phi_k = \int^D_S \left [ -E_k \left(\frac{dt}{dr}\right)_0 + p_r + J_k \left(\frac{d\phi}{dr}\right)_0 \right]dr,
\end{equation}
where the integral is taken along the light-ray trajectory. For simplicity, we have ignored the indices and used conservation of energy and angular momentum which give $ p_t^{(k)} = -E_k $, $ p_\phi^{(k)} = J_k $, where $J$ is the angular momentum of the test particle. Now from the equations for canonical momenta we can obtain
\begin{equation}
    \frac{dt}{dr} = \frac{-E_k}{p_r}\frac{B}{(-A)}, \ \ \ \ \frac{d\phi}{dr} = \frac{J_k}{p_r}\frac{B}{D}.
\end{equation}
And along the light-ray paths, these equations become
\begin{equation}\label{e-sup1}
    \left(\frac{dt}{dr}\right)_0 = \frac{E_0}{p_0(r)} \frac{B}{A}, \ \ \ \ \left(\frac{d\phi}{dr}\right)_0 = \frac{J_0}{p_0(r)}\frac{B}{D}.
\end{equation}
It is convenient to express the angular momentum $ J_k $ as a function of the energy $ E_k $, the impact parameter $ b $ and the velocity at infinity $ v_k^{(\infty)} $ as
\begin{equation}\label{e-sup2}
    J_k = E_k b v_k^{(\infty)}.
\end{equation}
Asymptotically, the metric is flat. So we can write 
\begin{equation}\label{e-sup3}
    \begin{aligned}
        v_k^{(\infty)} & = \frac{\sqrt{E_k^2-m_k^2}}{E_k} \simeq 1 - \frac{m_k^2}{2E_k^2}, \\
        J_k & \simeq E_k b \left( 1 - \frac{m_k^2}{2E_k^2} \right)\ ,
    \end{aligned}
\end{equation}
where in the last equality we used the relativistic approximation. For a massless particle, the angular momentum is 
\begin{equation}\label{e-sup4}
    J_0 = E_0 b.
\end{equation}
Using Eqs.~(\ref{e-sup1}), (\ref{e-sup2}), (\ref{e-sup3}) and (\ref{e-sup4}), the phase can be written as
\begin{equation}
    \Phi_k = \int^d_S \frac{E_0 E_k B}{p_0(r)}\left[ -\frac{1}{A} + \frac{p_0 p_k(r)}{E_0 E_k B} + \frac{ b^2 }{D}\left( 1 - \frac{m_k^2}{2E_k^2} \right) \right].
\end{equation}
Now from the mass-shell relation, we can find the following two relations
\begin{equation}\label{e-terms}
    \begin{aligned}
        \frac{p_0(r)}{E_0 B} & = \pm \sqrt{\frac{1}{A B} - \frac{b^2}{B D}}, \\
        \frac{p_0(r)p_k(r)}{E_0 E_k B} & = \frac{1}{A} - \frac{b^2}{D} - \frac{m_k^2}{2E_k^2}.
    \end{aligned}
\end{equation}
Using Eq.~(\ref{e-terms}) in the integral for the phase we can again write
\begin{equation}
    \Phi_k = -\int^D_S \frac{E_0 B}{p_0(r)} E_k \frac{m_k^2}{2E_k^2}dr.
\end{equation}
Using the second equation in relativistic approximation given in Eqs.~\eqref{rel-approx}, 
the phase becomes
\begin{equation}\label{e-phaseint}
    \begin{aligned}
        \Phi_k & = -\frac{m_k^2}{2E_0}\int^D_S \frac{E_0B}{p_0(r)}dr = \\
        & = \pm \frac{m_k^2}{2E_0}\int^D_S \sqrt{AB} \left( 1 - \frac{b^2A}{D} \right)^{-1/2} dr.
    \end{aligned}
\end{equation}
For $ b = 0$ and $ \gamma = 1 $, we recover the well-known result of the phase for a radially propagating neutrino in Schwarzschild spacetime
\begin{equation}
    \Phi_k \simeq \frac{m_k^2}{2E_0} |r_D - r_S|\ .
\end{equation}
Now we would like to integrate the integral in Eq.~(\ref{e-phaseint}) for two cases: (a) in the first, we shall consider the case where a neutrino is produced in a gravitational field and then propagates outward in the potential non-radially; (b) in the second, we shall consider the case where a neutrino is produced at a source $ S $, travels outward, moves around the massive object (which is effectively the gravitational lens), crossing the closest approach point at $ r = r_0 $ and reaches the detector $ D $. The second situation will be used for calculating the oscillation probability in a setting of gravitational lensing of neutrinos.  

Let us first look at case (a). In the weak field approximation, we expand the quantity under integral in Eq.~(\ref{e-phaseint}) as
\begin{equation}
    \Phi_k \simeq \pm \frac{m_k^2}{2E_0} \int^{r_D}_{r_S} \left[ \frac{1}{\sqrt{1- \frac{b^2}{r^2}}} + \frac{b^2(1-2\gamma)M/r}{(r^2-b^2)\sqrt{1-\frac{b^2}{r^2}}} \right] dr.
\end{equation}
Now this can be easily integrated and the result is 
\begin{equation}
    \begin{aligned}
        \Phi_k &= \frac{m_k^2}{2E_0}\Bigg[ \sqrt{r_D^2 - b^2} -\sqrt{r_S^2 - b^2}+ \\
        &+ (2\gamma - 1)M\left( \frac{r_D}{\sqrt{r_D^2 - b^2}} - \frac{r_S}{\sqrt{r_S^2 - b^2}} \right) \Bigg].
    \end{aligned}
\end{equation}
It can be seen from this expression that when $ b = 0 $, we recover the phase for the case of radial motion of neutrinos. 

In case (b) a propagating neutrino moves in the vicinity of the lensing object passing the closest point of approach at $ r=r_0 $. In this case, the integral for the phase can be written as the sum of two parts taking into account the sign of the momentum,
\begin{equation}\label{e-nonrad-phase-int}
    \begin{aligned}
        \Phi_k(r_S \rightarrow r_0 \rightarrow r_D) &= \frac{m_k^2}{2E_0}\int^{r_S}_{r_0} \sqrt{\frac{AB}{\left( 1 - \frac{b^2A}{D} \right)}} dr + \\
        &+ \frac{m_k^2}{2E_0}\int^{r_D}_{r_0} \sqrt{\frac{AB}{\left( 1 - \frac{b^2A}{D} \right)}}  dr. 
    \end{aligned}
\end{equation}

The position of closest approach of the neutrinos can be obtained from the equation
\begin{equation}
    \left( \frac{dr}{d\phi} \right)_0 = \frac{p_0(r_0)D}{J_0B} = 0.
\end{equation}
In the weak-field approximation, we solve this equation to obtain the point of closest approach as
\begin{equation}\label{e-r0gamma}
    r_0 \simeq b \left[ 1 - (2\gamma -1)\frac{M}{b} \right].
\end{equation}
Now integrating Eq.~\eqref{e-nonrad-phase-int} using the point of closest approach we get

\begin{equation}
    \begin{aligned}
        \Phi_k &= \frac{m_k^2}{2E_0} \Bigg[ \sqrt{r_D^2 - r_0^2} + \sqrt{r_S^2 - r_0^2}+ \\ 
        &+ (2\gamma -1)M
        \left(\sqrt{\frac{r_D - r_0}{r_D^2 + r_0^2}} + \sqrt{\frac{r_S - r_0}{r_S^2 + r_0^2}} \right) \Bigg],
    \end{aligned}
\end{equation}
or
\begin{equation}
    \begin{aligned}
        \Phi_k & \simeq \frac{m_k^2}{2E_0} \Bigg[ \sqrt{r_D^2 - b^2} + \sqrt{r_S^2 - b^2}+ \\
        &+ (2\gamma -1)M
        \Big( \frac{b}{\sqrt{r_S^2 - b^2}} + \frac{b}{\sqrt{r_D^2 - b^2}} +\\
        &+ \frac{r_S - b}{\sqrt{r_S + b}} ++ \frac{r_S - b}{\sqrt{r_S + b}} \Big) \Bigg] \ .
    \end{aligned}
\end{equation}
Now, for $ b << r_{S,D} $ expanding the above quantity over $ b/r_S $ and $ b/r_D $ and keeping terms of the order up to $ (b^2/r_{S,D}^2) $ we obtain 
\begin{equation}\label{e-phase-nr}
    \Phi_k = \frac{m_k^2}{2E_0}(r_S+r_D)\left[ 1 - \frac{b^2}{2r_Sr_D} + \frac{2(2\gamma - 1)M}{r_S+r_D} \right].
\end{equation}
As we can see, the phase now depends on $ \gamma $ in the first order in ($ M/r_{S,D} $). This is because, all the particles travelling non-radially will be affected by the deformed geometry. This expression will be used to discuss gravitational lensing of neutrinos in $ \gamma $-spacetime in the following section.

\begin{figure*}[t]
    \includegraphics[width=\textwidth]{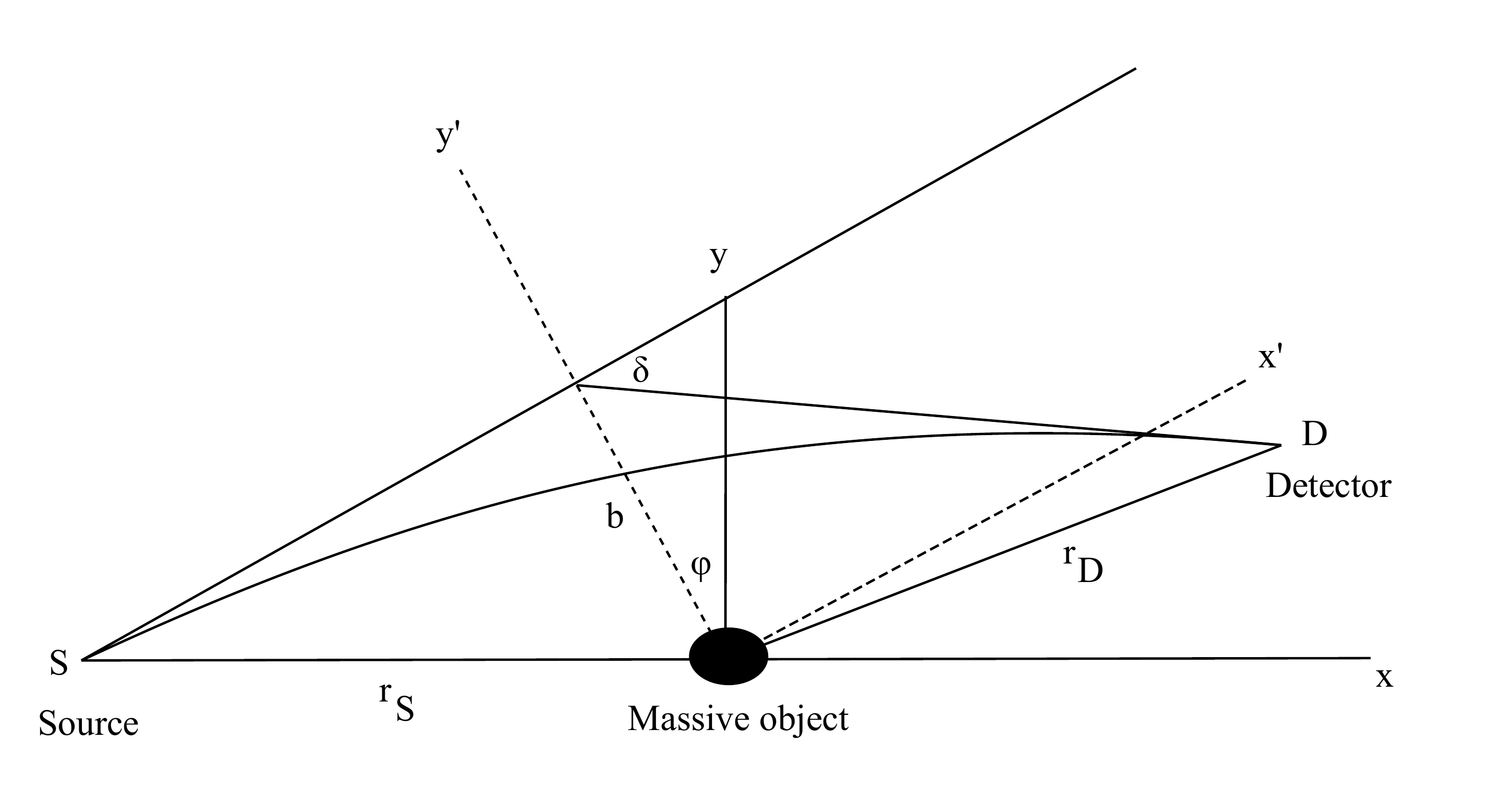}
    \caption{Schematic diagram for weak lensing of neutrinos in the $ \gamma $-spacetime. Neutrinos propagate from the source S to detector D in the exterior of a static and non spherical massive object which is described by the $\gamma$-metric. 
    }
    \label{drawing}
\end{figure*}

\begin{figure*}[t]
	\begin{center}
		\includegraphics[width=17cm]{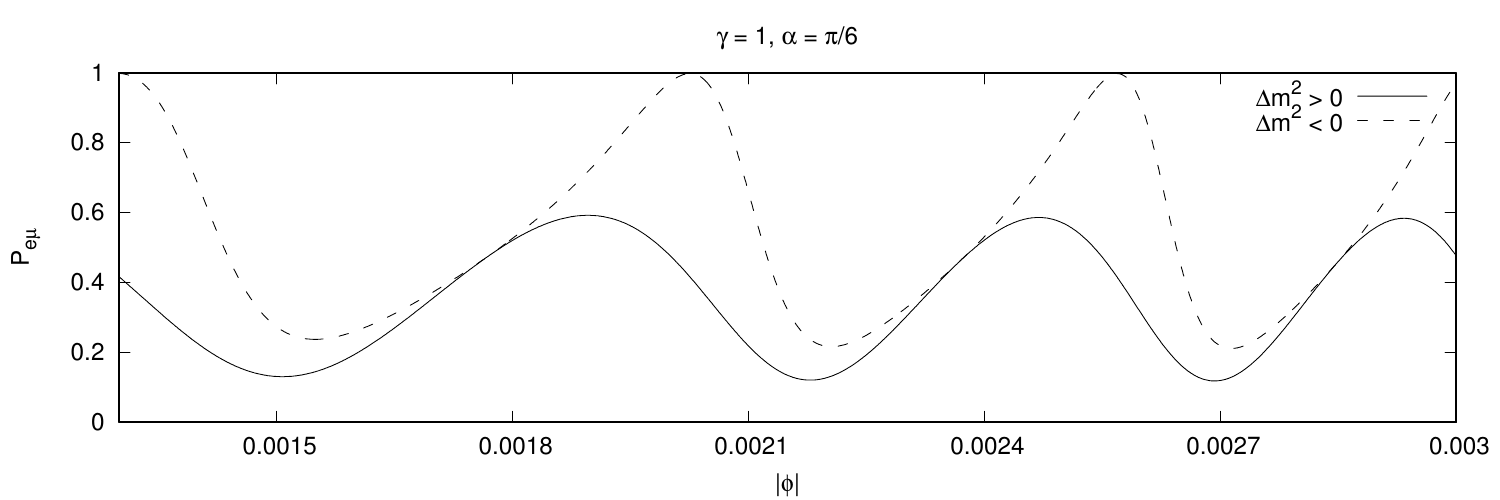}
		\includegraphics[width=17cm]{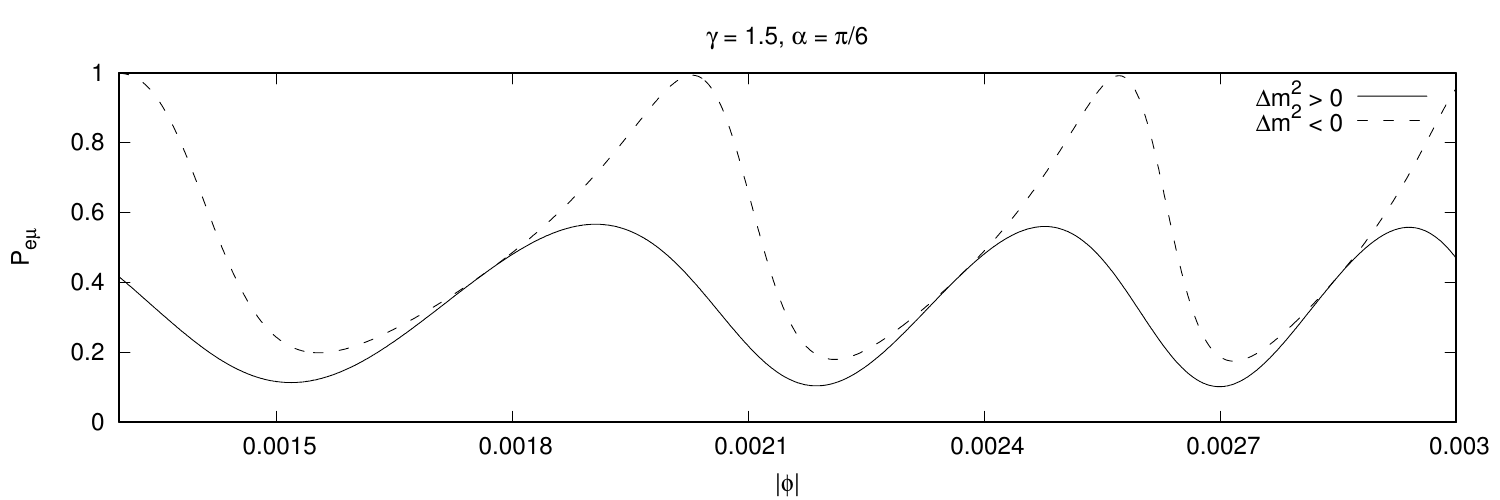}
	\end{center}
	\vspace{-0.5cm}
	\caption{ 
	Top panel: Neutrino oscillation probability for $ \gamma = 1.0 $ when the mixing angle is $ \alpha = \pi/6 $. Bottom panel: Neutrino oscillation probability for $ \gamma = 1.5 $ when the mixing angle is $ \alpha = \pi/6 $. The solid and the dashed curves represent normal hierarchy and inverted hierarchy respectively. Values of the other parameters are as follows: $ M_{\rm ADM} = 1 M_{\odot} $, $ \Delta m^2  = 10^{-3}\; {\rm eV}^2 $, and the lightest neutrino here is considered to be massless.     \label{fig2} }
\end{figure*}


\begin{figure*}[]
	\begin{center}
		\includegraphics[width=17cm]{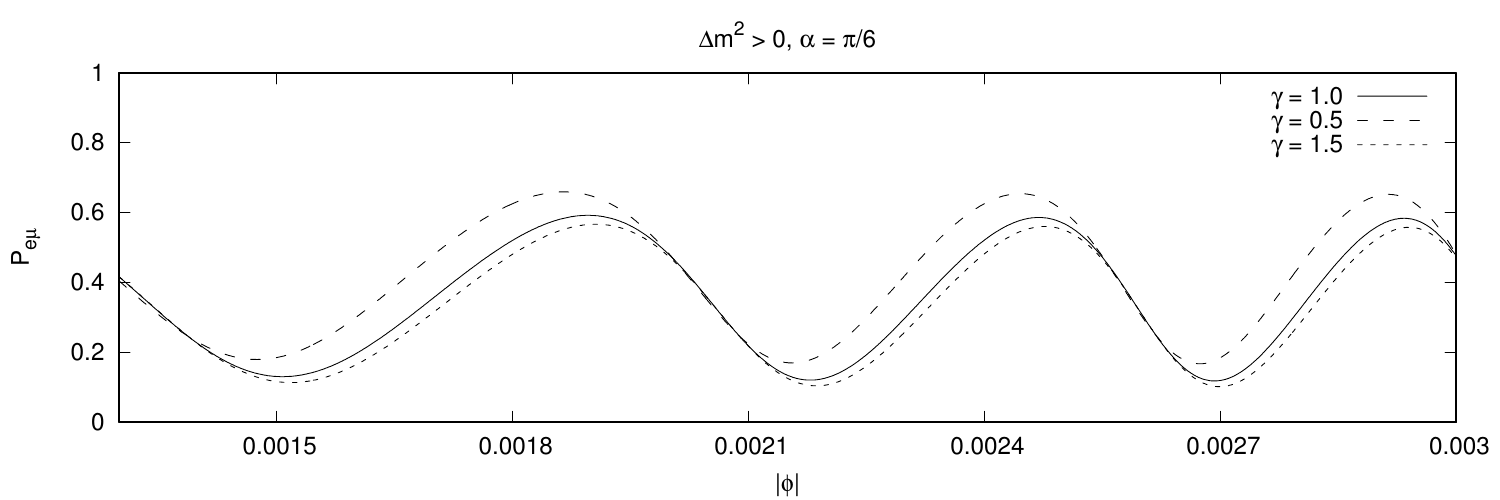}
		\includegraphics[width=17cm]{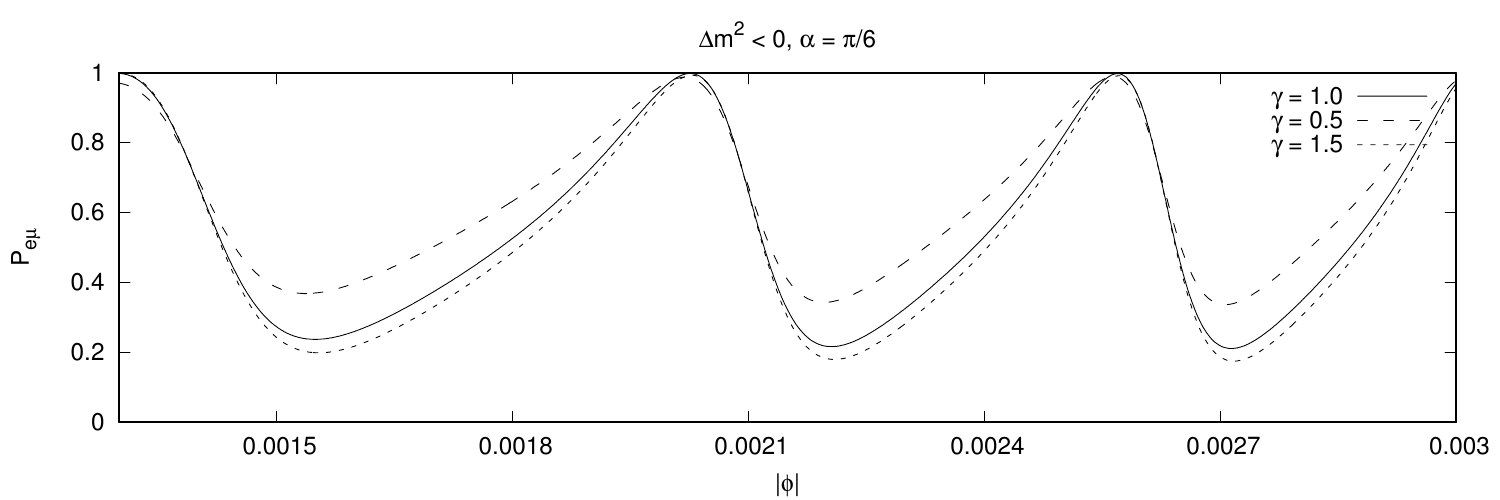}
	\end{center}
	\vspace{-0.5cm}
	\caption{Top panel: Neutrino oscillation probability as a function of azimuthal angle $\phi$ for $ \gamma = 1.0 $ (solid line),$ \gamma = 0.5 $ (dashed line) and $ \gamma = 1.5 $ (dotted line) for normal hierarchy $ \Delta m^2 > 0 $. Bottom panel: Neutrino oscillation probability for $ \gamma = 1.0 $ (solid line),$ \gamma = 0.5 $ (dashed line) and $ \gamma = 1.5 $ (dotted line) for inverted hierarchy $ \Delta m^2 < 0 $. The mixing angle here is $ \alpha = \pi/6 $. Values of the other parameters are as follows: $ M_{\rm ADM} = 1 M_{\odot} $, $ \Delta m^2  = 10^{-3}\; {\rm eV}^2 $, and the lightest neutrino here is considered to be massless.\label{fig3} }
\end{figure*}

\section{Neutrino oscillation probability}\label{sec:oscprob}

Let us now consider neutrinos with mass eigentaste $ \nu_i $ travelling in the $ \gamma $-spacetime, through different classical paths, from a source S to meet at a detector D. The neutrino flavor eigenstate $ \nu_\alpha $, propagated from the source to the detector through a path denoted by $ p $, is given by,
\begin{equation}
    \ket{\nu_\alpha(t_D,x_D)} = N\sum_i U^*_{\alpha i}\sum_p \exp \left( -i\Phi_i^p\ket{\nu_i(t_S,x_S)} \right).
\end{equation}
Here $ \Phi_i^p $ is given by Eq.~(\ref{e-phase-nr}) with $ b_p $ as the path dependent impact parameter. If a neutrino is produced in $ \alpha $ flavor eigenstate at the source $ S $ then the probability of it being detected in $ \beta $ flavor at the detector location, is given by \cite{Swami:2020qdi}
\begin{equation}\label{e-prob-pd}
    \begin{aligned}
        \mathcal{P}_{\alpha\beta}^{\rm lens} &= | \braket{\nu_\beta | \nu_\alpha(t_D,x_D)} |^2 = \\
        &= |N|^2\sum_{i,j}U_{\beta i}U_{\beta j}^*U_{\alpha j}U_{\alpha i}^*\sum_{p,q}\exp\left( -i\Delta\Phi_{ij}^{pq} \right),
    \end{aligned}
\end{equation}
where 
\begin{equation}
    |N|^2 = \left( \sum_{i}|U_{\alpha i}|^2 \sum_{p,q}\left( -i\Delta\Phi_{ii}^{pq} \right) \right)
\end{equation}
is the normalization constant and $ \Delta\Phi_{ij}^{pq} $ can be given as 
\begin{equation}\label{eq-phasediff}
    \begin{aligned}
        \Delta\Phi_{ij}^{pq} &= \Phi_p^i - \Phi_q^j =  \Delta m_{ij}^2 {\rm A}_{pq} + \Delta b_{pq}^2 {\rm B}_{ij},
    \end{aligned}
\end{equation}
where 
\begin{equation}\label{eq-aijbijcij}
    \begin{aligned}
        {\rm A}_{pq} &= \frac{r_S+r_D}{2E_0}\Bigg( 1 + \frac{2(2\gamma-1)M}{2r_Sr_D} - \frac{\Sigma b_{pq}^2}{4r_Sr_D}\Bigg), \\
        {\rm B}_{ij} &= -\frac{\Sigma m_{ij}^2}{8E_0}\left( \frac{1}{r_S} + \frac{1}{r_D} \right).
    \end{aligned}
\end{equation}
In the above equations, the quantities $ \Delta m_{ij}^2 $, $ \sum m_{ij}^2 $, $ \Delta b_{pq}^2 $ and $ \sum b_{pq}^2 $ are given by
\begin{equation}
    \begin{aligned}
        & \Delta m_{ij}^2 = m_i^2 - m_j^2, \ \ \ \sum m_{ij}^2 = m_i^2 + m_j^2, \\
        & \Delta b_{pq}^2 = b_p^2 - b_q^2, \ \ \ \sum b_{pq}^2 = b_p^2 + b_q^2, 
    \end{aligned}
\end{equation}
respectively. Based upon these, we can make the following observations.
\begin{itemize}
    \item $ {\rm A}_{pq} $ and $ {\rm B}_{ij} $ are symmetric under interchange of their indices, $ {\rm A}_{pq} = {\rm A}_{qp} $ and $ {\rm B}_{ij} = {\rm B}_{ji} $. 
    \item The oscillation probability depends on the sum of individual mass squared of neutrinos $ \sum m_{ij}^2 $ through $ \Delta b_{pq}^2 $.
    \item For those paths for which $ \Delta b_{pq}^2 $ vanishes, $ \mathcal{P}_{\alpha\beta}^{\rm lens} $ is invariant under the symmetry $ m_i^2 \rightarrow m_i^2 + C $. 
    \item For the paths for which $ \Delta b_{pq}^2 $ does not vanish, the symmetry $ m_i^2 \rightarrow m_i^2 + C $ is broken. The shift implies $ {\rm B}_{ij} \rightarrow {\rm B}_{ij} +2C $. 
    Therefore, a generic configuration of source S, lens and the detector D will retain the information about $ \sum m_{ij}^2 $. Hence the observation of lensed neutrinos would carry information about their absolute masses \cite{Swami:2020qdi}.
\end{itemize}

\section{Gravitational lensing of neutrinos in $ \gamma $-spacetime}
\label{sec:lensing}

We can see clearly from the expression for phase difference derived in previous section that the probability of neutrino oscillation depends on $ \Sigma m_{ij}^2 $ through a path dependent term $ \Delta b_{pq}^2 $. This was also shown in two and three flavour neutrino oscillation case in \cite{Swami:2020qdi} for Schwarzschild. Similarly, in the $ \gamma $-spacetime, it appears that the absolute neutrino mass dependent effect is same as in Schwarzschild, since $ {\rm B}_{ij} $ does not depend on $ \gamma $. However, we shall see later that the impact parameters $ b_p $ depend on $ \gamma $ and hence we should be able to see the effect of the quadruple deformation parameter $ \gamma $ in both terms of the phase difference \eqref{eq-phasediff}.

Now we can simplify the expression for the oscillation probability in order to understand the effect of $ \gamma $ more clearly. For simplicity, first we shall consider a toy model with two neutrino flavors. Specifically, we shall evaluate the two flavor neutrino oscillation probability at a generic point in a plane connecting the source, the lens and the detector in the weak field limit. 
Therefore we substitute equation \eqref{eq-phasediff} in equation \eqref{e-prob-pd} to get
\begin{widetext}
    \begin{equation}\label{eq-P-master}
    \mathcal{P}^{\rm lens}_{\alpha\beta} = |N|^{2}\left[\sum_{i,j} U_{\beta i} U^*_{\beta j} U_{\alpha j} U^*_{\alpha i} \Bigg( \sum_{p = q}  \exp\left(-i \Delta m^2_{ij}{\rm A}_{pp} \right) + 2\sum_{p> q}\cos\left( \Delta b_{pq}^2\, {\rm B}_{ij}\right)\exp\left(-i \Delta m^2_{ij}\, {\rm A}_{pq}\right)\Bigg)\right]\ ,
    \end{equation}
\end{widetext}
where $ |N|^2 $ is given by 
\begin{equation}
|N|^2 = \left(N_{\rm path} + \sum_i |U_{\alpha i}|^2 \sum_{q>p} 2\cos(\Delta b_{pq}^2{\rm B}_{ii})\right)^{-1}.
\end{equation}
Here, for simplicity, we consider neutrino propagation in the equatorial ($ \theta = \pi/2 $) plane and in this case,
$ N_{\rm path} = 2 $. Now the general expression of probability reduces to a simpler form
\begin{equation}\label{e-prob-gen}
    \begin{aligned}
        \mathcal{P}_{\alpha\beta}^{\rm lens} &= |N|^2 \Bigg[ 2\sum_i |U_{\beta i}|^2|U_{\alpha i}|^2 (1 + 
        \cos(\Delta b^2 {\rm B}_{ii}))+  \\
        &+ \sum_{i,j\neq i} U_{\beta i}U_{\beta j}^* U_{\alpha j} U_{\alpha i}^* \big[\exp(-i\Delta m_{ij}^2 {\rm A}_{11})+ \\
        &+ \exp(-i\Delta m_{ij}^2 {\rm A}_{22})+ \\
        &+ 2\cos(\Delta b^2 {\rm B}_{ij}) \exp(-i\Delta m_{ij}^2 {\rm A}_{12}) \big] \Bigg]\ ,
    \end{aligned}
\end{equation}
with $ {\rm A}_{pq} $ and $ {\rm B}_{ij} $ being given by equation \eqref{eq-aijbijcij} and the normalization factor reduces to
\begin{equation}
    |N|^2 = \left( 2 + 2 \sum_i |U_{\alpha i}|^2 \cos(\Delta b^2 {\rm B}_{ii}) \right)^{-1}\ ,
\end{equation}
where $ \Delta b^2 = \Delta b_{12}^2 $.
Now for the qualitative  and quantitative understanding of neutrino oscillation in the $ \gamma $-spacetime, we consider the simplest two neutrino flavors oscillation, $ \nu_e \rightarrow \nu_\mu $. In this case, the oscillation probability becomes
\begin{equation}
    \begin{aligned}
        \mathcal{P}_{e\mu}^{\rm lens} &= |N|^2 \sin^22\alpha \Bigg[ \sin^2\left( \Delta m^2 \frac{{\rm A}_{11}}{2} \right) + \sin^2\left( \Delta m^2 \frac{{\rm A}_{22}}{2} \right)+ \\
        &- \cos(\Delta b^2 {\rm B}_{12})\cos (\Delta m^2 {\rm A}_{12}) + \frac{1}{2}\cos (\Delta b^2 {\rm B}_{11})+ \\
        &+ \frac{1}{2} \cos (\Delta b^2 {\rm B}_{22}) \Bigg],
    \end{aligned}
\end{equation}
where
\begin{equation}
    |N|^2 = \left[ 2(1 + \cos^2\alpha \cos(\Delta b^2 {\rm B}_{11}) \sin^2\alpha \cos(\Delta b^2 {\rm B}_{22}) ) \right]^{-1},
\end{equation}
and $ \Delta m^2 = \Delta m_{21}^2 $.

\subsection{Two flavor case: numerical results}

\begin{figure*}[]
	\begin{center}
		\includegraphics[width=8cm]{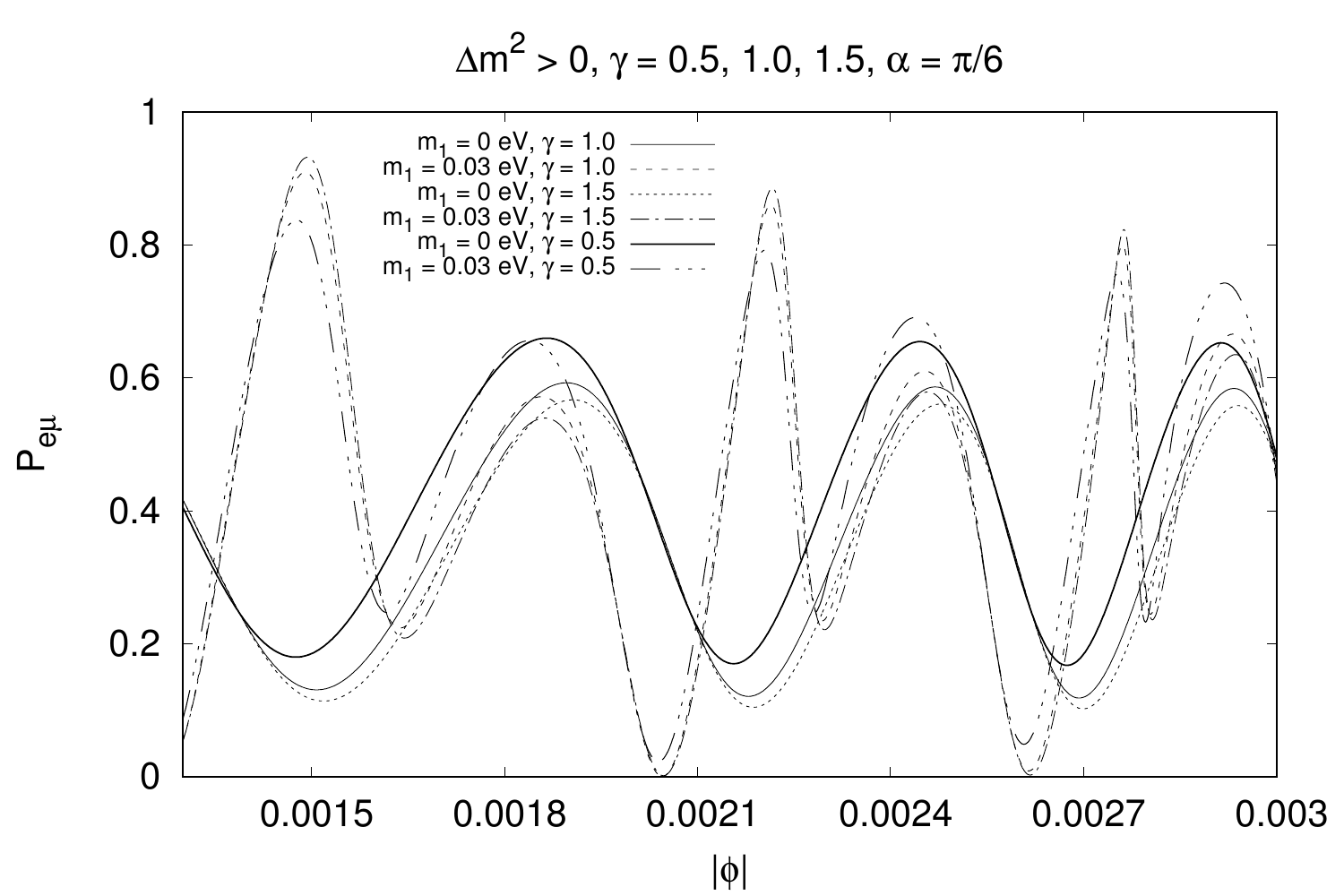}
		\includegraphics[width=8cm]{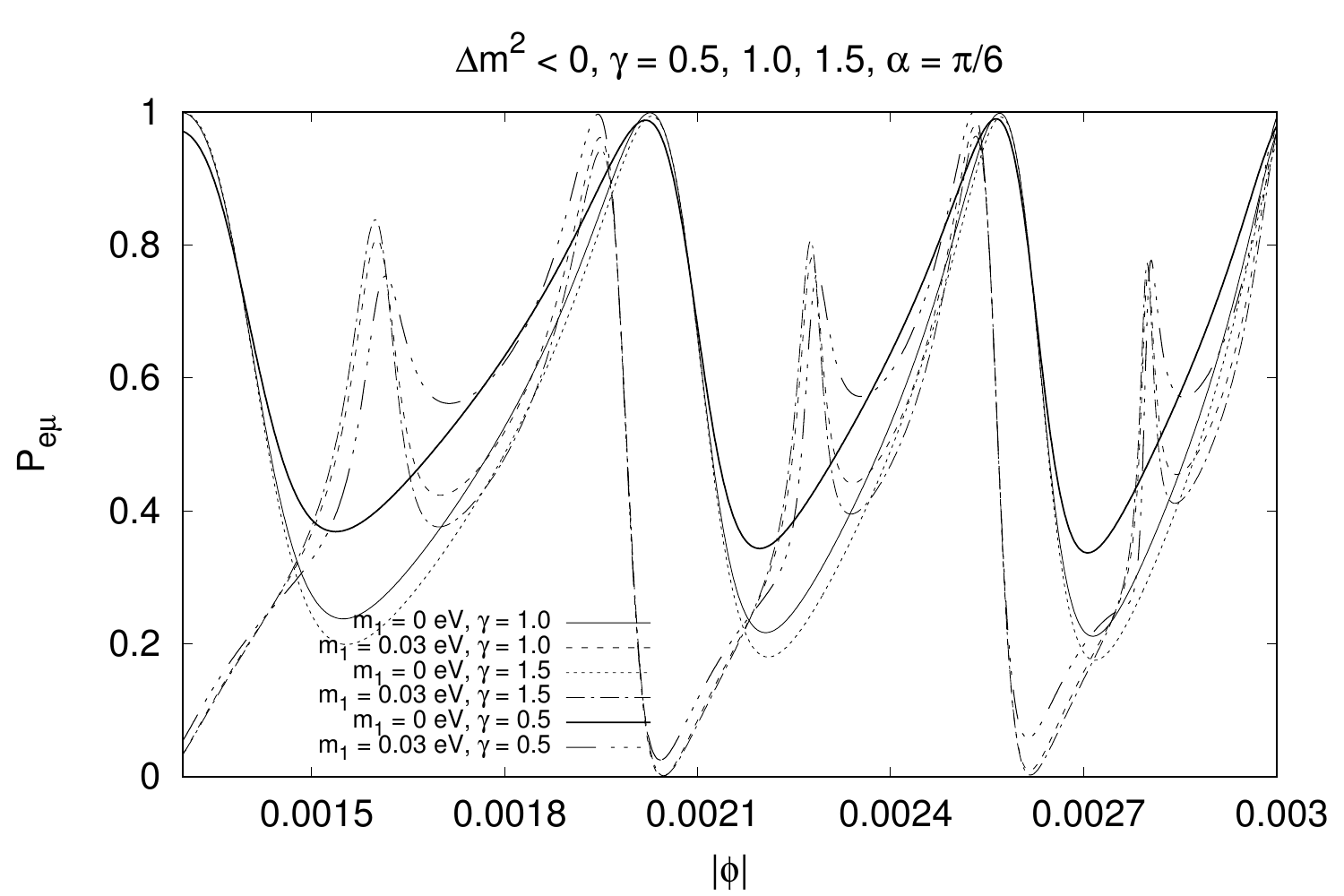}
	\end{center}
	\vspace{-0.5cm}
	\caption{Left panel: Neutrino oscillation probability as a function of azimuthal angle $\phi$ for $ \gamma = 1.0 $, $ 0.5 $ and $ 1.5 $ with $ \Delta m^2 > 0 $ (normal hierarchy). Here the solid line represents $ m_1 = 0 $ eV and $ \gamma = 1.0 $, the dashed line represents $ m_1 = 0.03 $ eV and $ \gamma = 1.0 $, the dotted line represents $ m_1 = 0 $ eV and $ \gamma = 1.5 $, the dotted-dashed line represents $ m_1 = 0.03 $ eV and $ \gamma = 1.5 $, the thick solid line represents $ m_1 = 0 $ eV and $ \gamma = 0.5 $, and the triple-dotted dashed line represents $ m_1 = 0.03 $ eV and $ \gamma = 0.5 $. Right panel: Neutrino oscillation probability as a function of azimuthal angle $\phi$ for $ \gamma = 1.0 $, $ 0.5 $ and $ 1.5 $ with $ \Delta m^2 < 0 $ (inverted hierarchy). Here the solid line represents $ m_1 = 0 $ eV and $ \gamma = 1.0 $, the dashed line represents $ m_1 = 0.03 $ eV and $ \gamma = 1.0 $, the dotted line represents $ m_1 = 0 $ eV and $ \gamma = 1.5 $, the dotted-dashed line represents $ m_1 = 0.03 $ eV and $ \gamma = 1.5 $, the thick solid line represents $ m_1 = 0 $ eV and $ \gamma = 0.5 $, and the triple-dotted dashed line represents $ m_1 = 0.03 $ eV and $ \gamma = 0.5 $. The other parameters are as follows: $ M_{\rm ADM} = 1 M_{\odot} $, $ \Delta m^2  = 10^{-3} {\rm eV}^2 $. \label{fig4} }
\end{figure*}

\begin{figure*}[]
	\begin{center}
		\includegraphics[width=8cm]{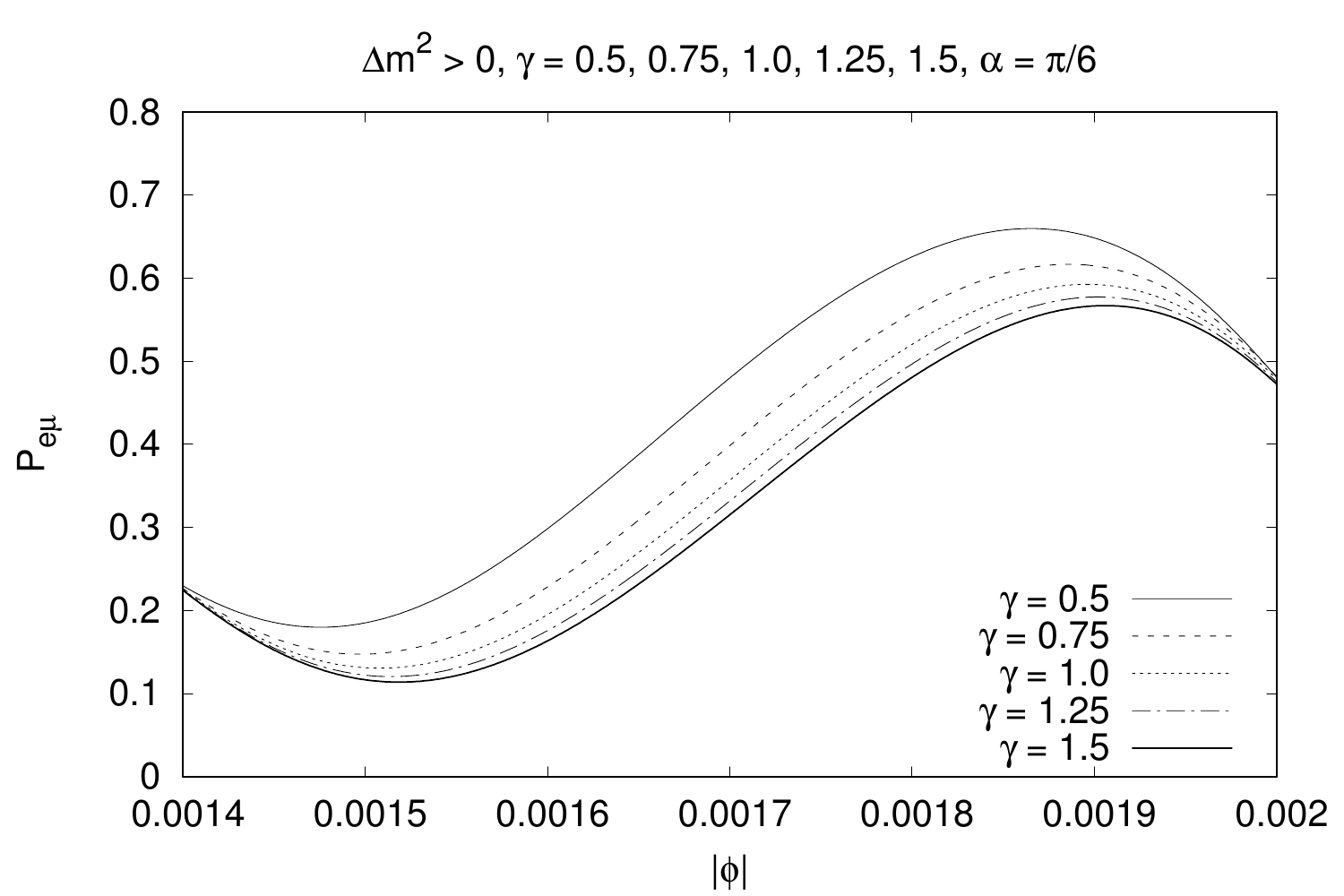}
		\includegraphics[width=8cm]{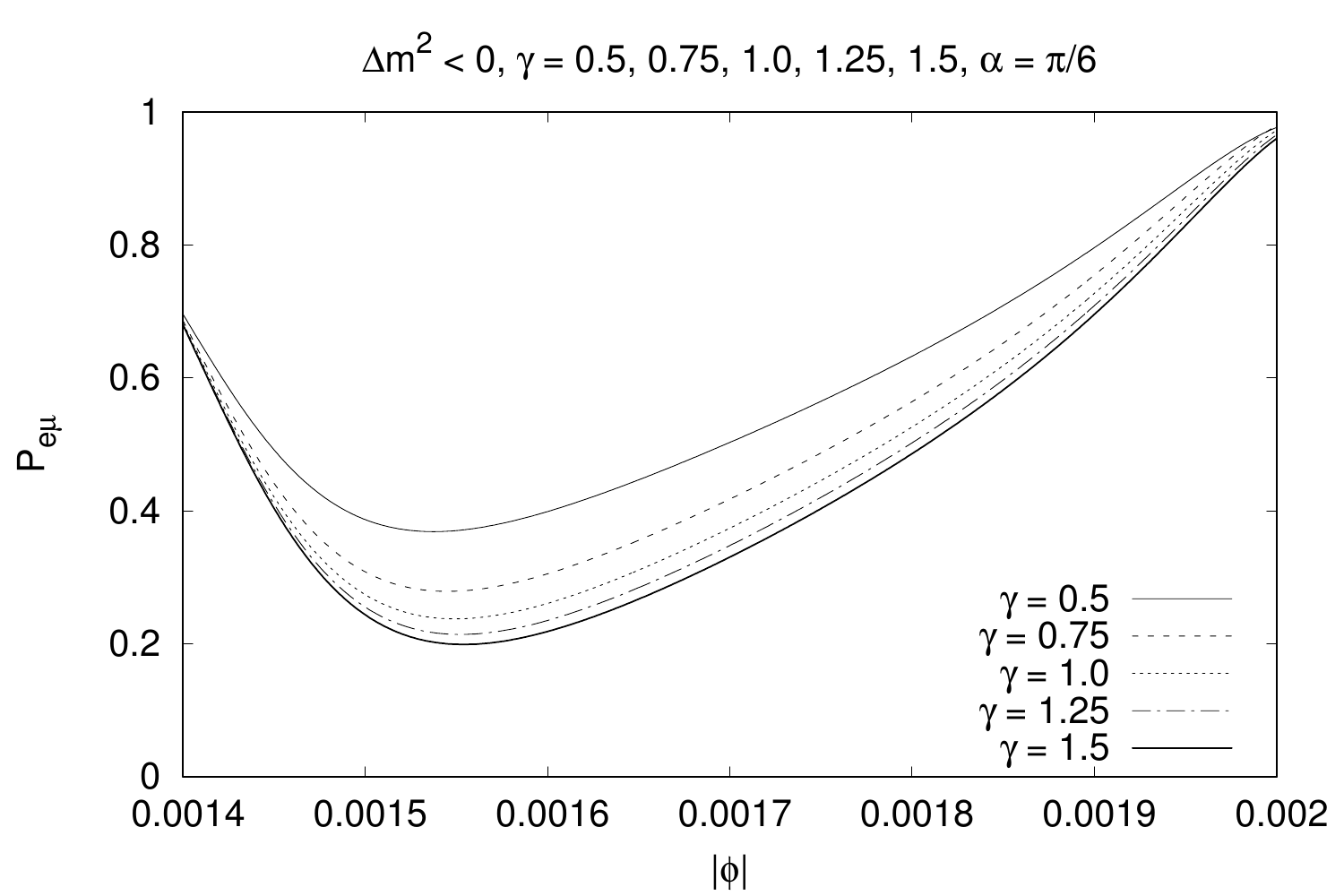}
	\end{center}
	\vspace{-0.5cm}
	\caption{Zoomed-in version of the $ \gamma $ dependence for the neutrino oscillation probability. Left panel: Neutrino oscillation probability as a function of azimuthal angle $\phi$ for $ \gamma = 0.5 $ (thin solid line), $ \gamma = 0.75 $ (dashed line), $ \gamma = 1.0 $ (dotted line), $ \gamma = 1.25 $ (dotted-dashed line), $ \gamma = 1.5 $ (thick solid line) in case of normal hierarchy. Right panel: Neutrino oscillation probability for $ \gamma = 0.5 $ (thin solid line), $ \gamma = 0.75 $ (dashed line), $ \gamma = 1.0 $ (dotted line), $ \gamma = 1.25 $ (dotted-dashed line), $ \gamma = 1.5 $ (thick solid line) in case of inverted hierarchy.  \label{fig5}}
\end{figure*}

Now for a quantitative understanding, we would like to see how the probability changes over some change in the lensing parameters. Therefore it is useful express the impact parameter $ b_p $ in terms of the geometric quantities of the system. To this aim we can refer to Figure \ref{drawing}, which shows a schematic representation of a weak lensing event in $ \gamma $-spacetime. Here we have the source (S) of neutrinos, the gravitational lens characterized by the $ \gamma $-metric and the detector (D) on a plane. Physical distances from the source to the lens and from the lens to the detector are $ r_S $ and $ r_D $, respectively, in the $ (x,y) $ coordinate system. We can also consider another coordinate system $ (x',y') $ obtained by rotating the original coordinate system $ (x,y) $ by an angle $ \varphi $ such that $ x' = x \cos \varphi + y \sin \varphi $ and $  y' = -x \sin \varphi + y \cos \varphi $. In the rotated frame, the angle of deflection of the neutrino, $ \delta $ from its original path is related to the impact parameter $ b $ by the following relation
\begin{equation}\label{eq-geom}
    \delta \sim \frac{y_D' - b}{x_D'} = - \frac{4\gamma M}{b} = -\frac{2R_x}{b},
\end{equation}
where $ (x_D',y_D') $ is the location of the detector, $ R_x = 2\gamma M $ and we have used the expression for $ \delta $ in the weak lensing limit. Now using the identity $ \sin \varphi = b/r_S $ from Figure~\ref{drawing}, the previous equation Eq.~\eqref{eq-geom} can be written as
\begin{equation}\label{eq-quart}
    (2R_x x_D + by_D)\sqrt{1 - \frac{b^2}{r_S^2}} = b^2 \left( \frac{x_D}{r_S} +1 \right) - \frac{2R_x b y_D}{r_S}. 
\end{equation}
Solution of this equation gives the impact parameters in terms of $ r_S $, $ R_x $ and the lensing location $ (x_D, y_D) $. As a numerical exercise we can consider the Sun-Earth system with typical values of the geometrical quantities and assume that the gravitational field of the Sun is represented by the $ \gamma $-metric while the Earth is taken to be the detector. The source is considered to be situated behind the Sun and it emits relativistic neutrinos with typical energy $ E_0 = 10 \ \text{MeV} $. Now assuming a circular trajectory of the detector around the Sun such that $ x_D = r_D \cos \phi $ and $ y_D = r_D \sin \phi $, we can numerically solve the quartic polynomial given by Eq.~\eqref{eq-quart} in the equatorial plane and obtain two positive real roots $ b_1 $ and $ b_2 $ for every $ \phi $. In this system, $ r_D = 10^8 \ \text{km} $ and $ r_S = 10^5 r_D $. 
In this numerical exercise, neutrino oscillation probability is calculated only for those value of $ b_p $ for which $ R_x \ll b_p \ll r_D $. Other relevant parameters are $ M_{\rm ADM} = \gamma M = 1 M_\odot $ (as defined in \eqref{eq:ADM}) and $ |\Delta m^2| = 10^{-3} \; {\rm eV}^2 $. Note that, these numbers are for illustrative purposes only and in a realistic scenario proper numerical values of the geometric parameters of the model have to be selected.

Oscillation probabilities for the two-flavor toy model of neutrinos are shown in Figure \ref{fig2}, \ref{fig3}, \ref{fig4}, and \ref{fig5} as a function of the azimuthal angle $\phi$. Note that our main aim is to investigate the dependence of the oscillation probability on the deformation parameter $ \gamma $. Figure \ref{fig2} shows the neutrino oscillation probability $ \nu_e \rightarrow \nu_\mu $ for $ \gamma = 1.0 $ (top panel) and $ \gamma = 1.5 $ (bottom panel)\footnote{It is worth mentioning that the actual realistic values of the $\gamma$ parameter are expected to be much closer to $1$ both for the solar system as well as for any astrophysical compact object. Here, we choose large values of $\gamma$ for illustrative purposes only.}. In each of these two plots shown in Figure \ref{fig2}, the solid line corresponds to normal hierarchy ($ \Delta m^2 > 0 $)  and the dashed line corresponds to inverted hierarchy ($ \Delta m^2 < 0 $). The mixing angle here is $ \alpha = \pi/6 $. The first thing we notice from these three plots is that the oscillation probability is different for normal and inverted mass ordering (except for some values of $ \phi $) confirming the absolute mass dependent effects in gravitational lensing \cite{Swami:2020qdi}. This is in sharp contrast with two flavour vacuum oscillation probability in flat spacetime where the sign of $\Delta m^2$ does not make any distinction. We also notice an increase in frequency for higher values of $ \phi $. 
Figure \ref{fig3} shows the same results in order to better highlight the dependence of the probability on $\gamma$.

The top panel in Figure \ref{fig3} shows the conversion probability for $ \gamma = 1 $ (solid line), $ \gamma = 0.5  $ (dashed line) and $ \gamma = 1.5 $ (dotted line) for normal hierarchy ($ \Delta m^2 > 0 $). The bottom panel shows the conversion probability for $ \gamma = 1 $ (solid line), $ \gamma = 0.5 $ (dashed line) and $ \gamma = 1.5 $ (dotted line) for inverted hierarchy ($ \Delta m^2 < 0 $). It can be clearly seen from the plots that the conversion probability is different for different values of the $ \gamma $ parameter even though the observable mass $ M_{\rm ADM} = \gamma M $ is same. In standard optical observations, such as deflection of light in weak-field limit, it is not possible to constrain the parameter $ \gamma $ independently as it appears together with the mass parameter $ M $, creating a degeneracy. For example, the deflection angle in the weak-field limit is $ \delta  = -4\gamma M/b =-4M_{\rm ADM}/b$ which remains unchanged if we change $\gamma$ and $M$ is such a way that $M_{\rm ADM}$ remains the same. However, this degeneracy is broken in the phase difference of neutrino oscillation and appears in $ {\rm A}_{pq} $ as $ 2(2\gamma-1)M/2r_Sr_D $ in the coefficients of neutrino mass squared difference ($ \Delta m_{ij}^2 $). Note that, $ {\rm B}_{ij} $ is independent of $ \gamma $ however, $ b_{p,q} $ and hence $ \Delta b_{pq}^2 $ and $ \Sigma b_{pq}^2 $ are $ \gamma $ dependent as the paths will be affected by the deviations from the spherical symmetry (see Eq.~\eqref{eq-quart}). Therefore, the deviations of the geometry from spherical symmetry can in principle be constrained from neutrino observations. For values of $ \gamma < 1 $, we see a larger change in conversion probability than for values of $ \gamma > 1 $. 
With decrease in $ \gamma $, the conversion probability increases for both normal and inverted hierarchy as seen from Figure~\ref{fig5}. Also it is important to note that in Figures \ref{fig2}, \ref{fig3} and \ref{fig5}, the lightest neutrino is assumed to be massless.


In Figure \ref{fig4}, we compare the conversion probability of the massless case for lightest neutrino with the massive case. The left panel in Figure \ref{fig4} shows conversion probabilities for the case of normal hierarchy for $ \gamma = 0.5, 1 \text{ and}, 1.5 $ when the lightest neutrino mass $ m_1 = 0 \ {\rm eV} $ and $ 0.03 \ {\rm eV} $. Similarly the left panel of Figure \ref{fig4} shows the case of inverted hierarchy for $ \gamma = 0.5,  1 \text{ and } 1.5 $ when the lightest neutrino mass is $ m_1 = 0 \ {\rm eV} $ and $ 0.03 \ {\rm eV} $.
We notice similar effects of $ \gamma $ on the conversion probability in this case as well. This, in principle, confirms that neutrino probability observation can constrain deviations from spherical symmetry of a gravitational system.

\begin{figure*}[]
    \begin{center}
        \includegraphics[width=17cm]{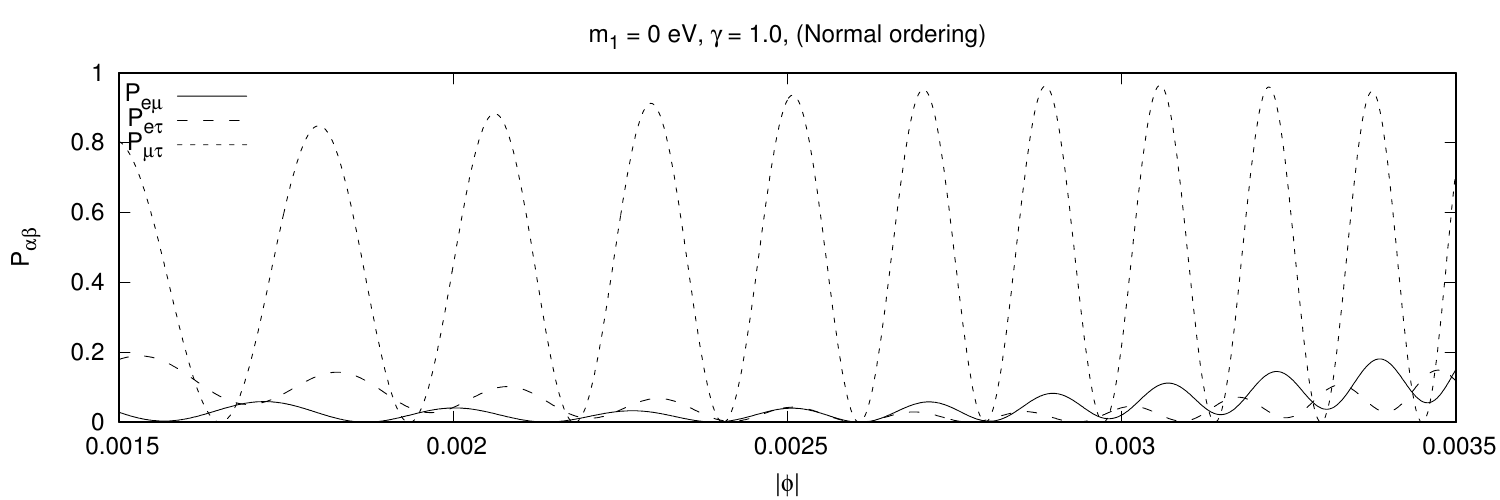}
    \end{center}
    \vspace{-0.5cm}
    \caption{Probability of $ \nu_\alpha \rightarrow \nu_\beta $ conversion, for different $ \alpha $ and $ \beta $, as function of the azimuthal angle $ \phi $ in the three flavor case for $ \gamma = 1 $ (spherical symmetry). The solid, dashed and the dotted lines represent the $ \nu_e \rightarrow \nu_\mu $, $ \nu_e \rightarrow \nu_\tau $ and $ \nu_\mu \rightarrow \nu_\tau $ conversion probabilities, respectively. We take $ r_D = 10^8 $ km, $ r_S = 10^5r_D $, $ M_{\rm ADM} = 1 M_\odot $ and $ E_0 = 10 $ MeV. Neutrino mass squared differences, mixing angles and the Dirac CP phase are taken from the latest global fit \cite{Esteban:2020cvm}.  \label{fig7}}
\end{figure*}

\begin{figure*}[t]
	\begin{center}
		\includegraphics[width=17cm]{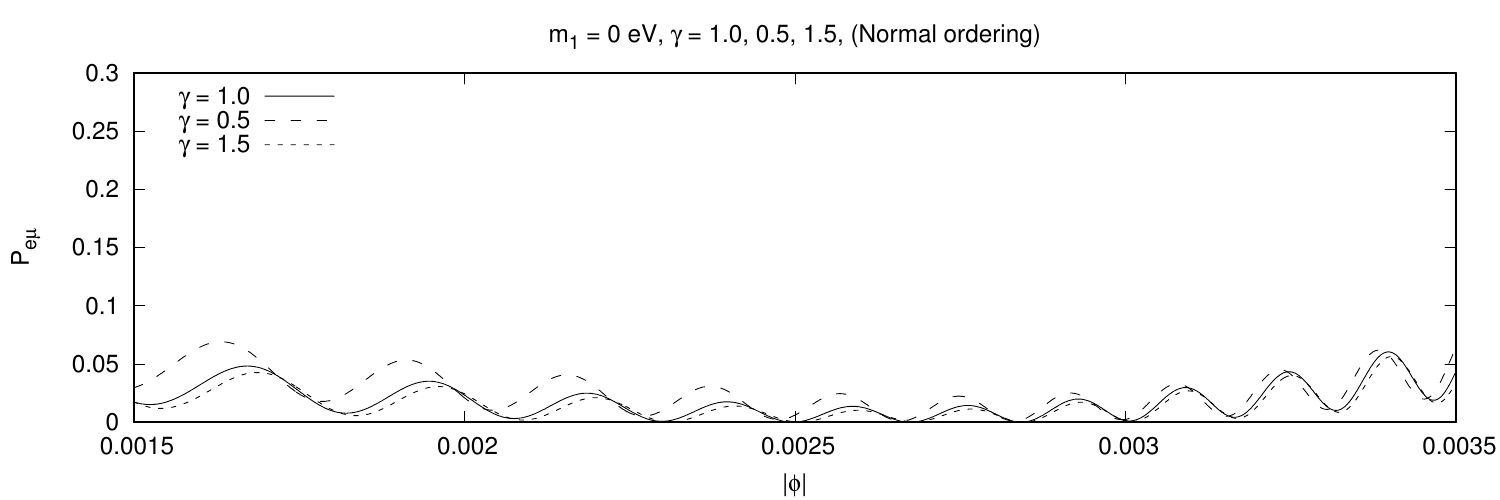}
		\includegraphics[width=17cm]{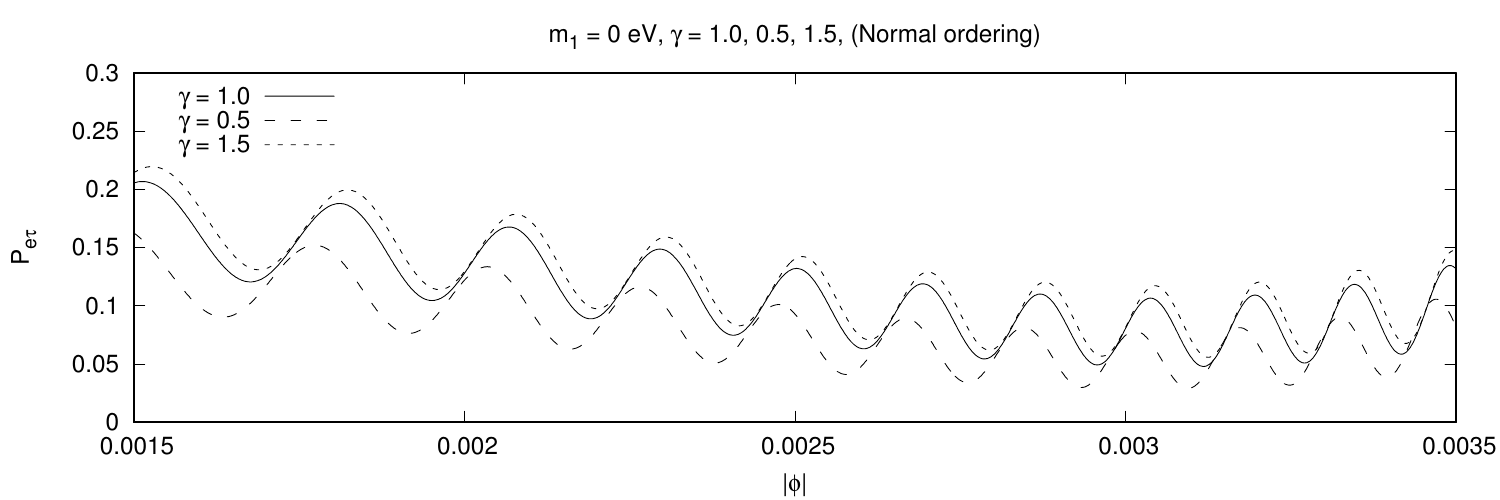}
		\includegraphics[width=17cm]{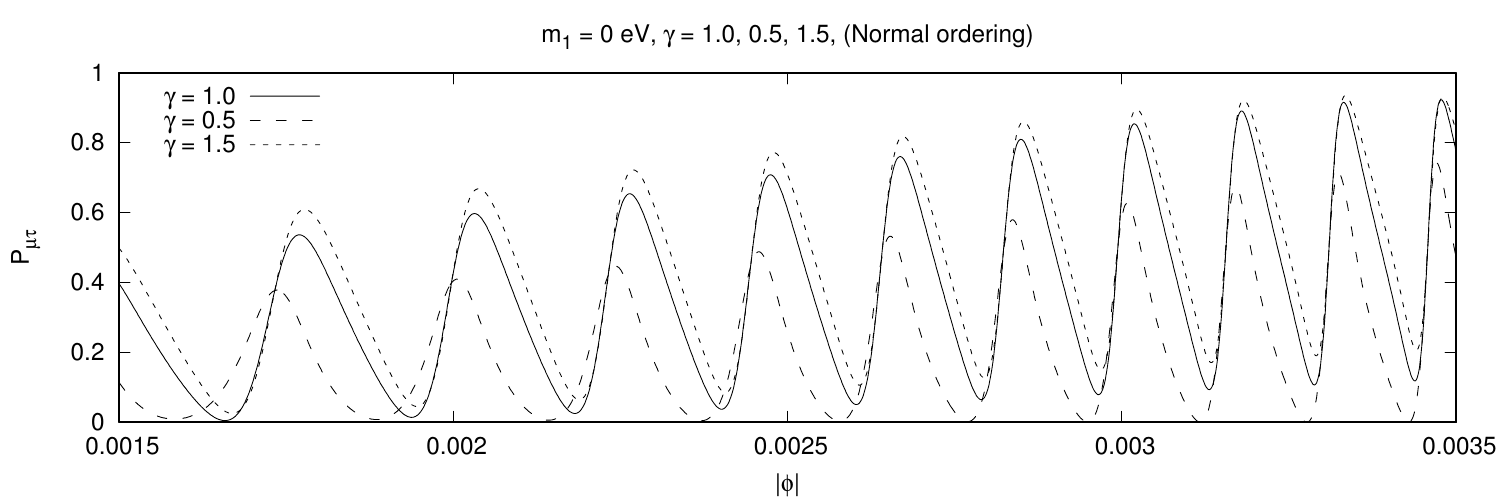}
	\end{center}
	\vspace{-0.5cm}
	\caption{Oscillation probability for the three flavor case in normal hierarchy with $ m_1 = 0 $. Top panel: Probability of $ \nu_e \rightarrow \nu_\mu $ conversion for $ \gamma = 1.0 $ (solid line), $ \gamma = 0.5 $ (dashed line), $ \gamma = 1.5 $ (dotted line). Mid panel: Probability of $ \nu_e \rightarrow \nu_\tau $ conversion for $ \gamma = 1.0 $ (solid line), $ \gamma = 0.5 $ (dashed line), $ \gamma = 1.5 $ (dotted line). Bottom panel: Probability of $ \nu_\mu \rightarrow \nu_\tau $ conversion for $ \gamma = 1.0 $ (solid line), $ \gamma = 0.5 $ (dashed line), $ \gamma = 1.5 $ (dotted line). We take $ r_D = 10^8 $ km, $ r_S = 10^5r_D $, $ M_{\rm ADM} = 1 M_\odot $ and $ E_0 = 10 $ MeV. Neutrino mass squared differences, mixing angles and the Dirac CP phase are taken from the latest global fit \cite{Esteban:2020cvm}. \label{fig8} }
\end{figure*}

\begin{figure*}[t]
    \begin{center}
		\includegraphics[width=17cm]{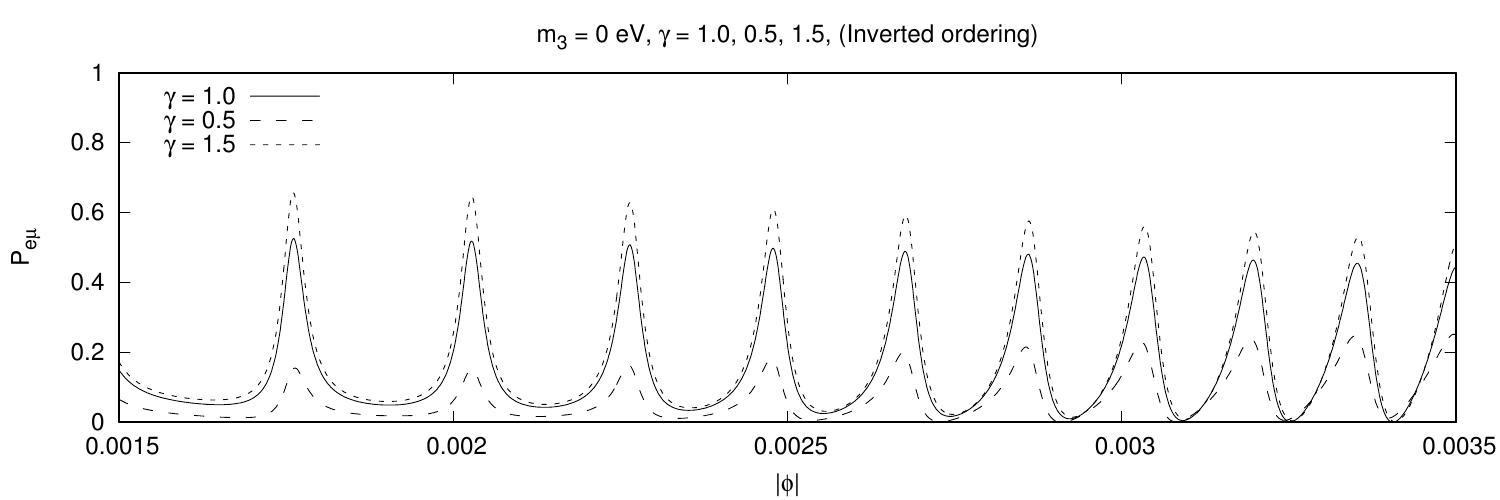}
		\includegraphics[width=17cm]{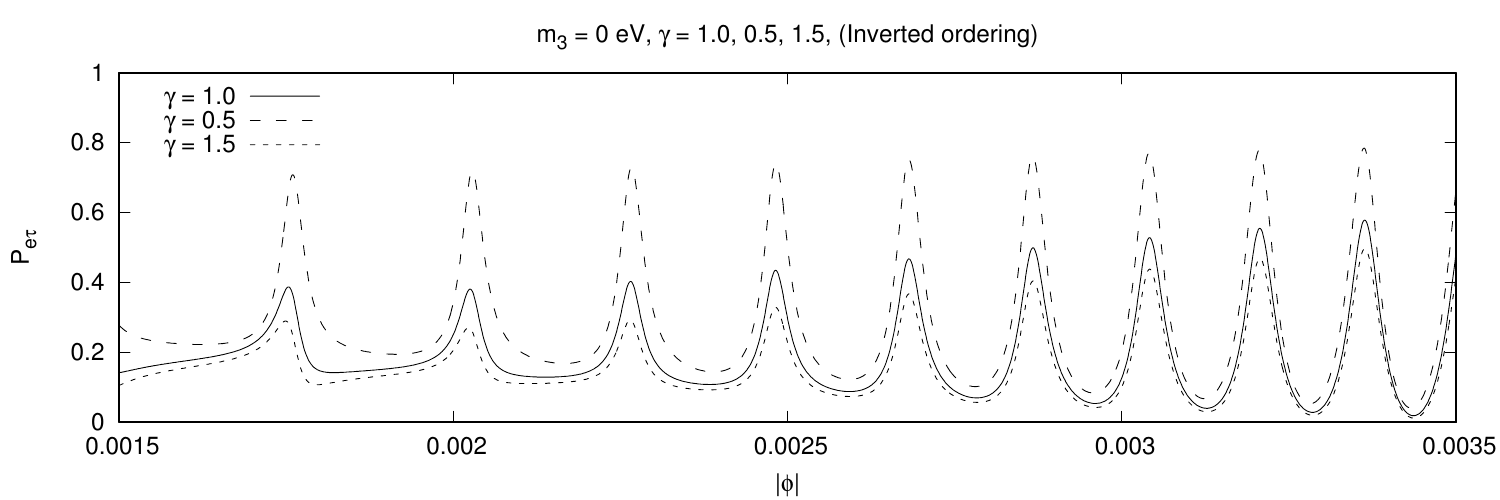}
		\includegraphics[width=17cm]{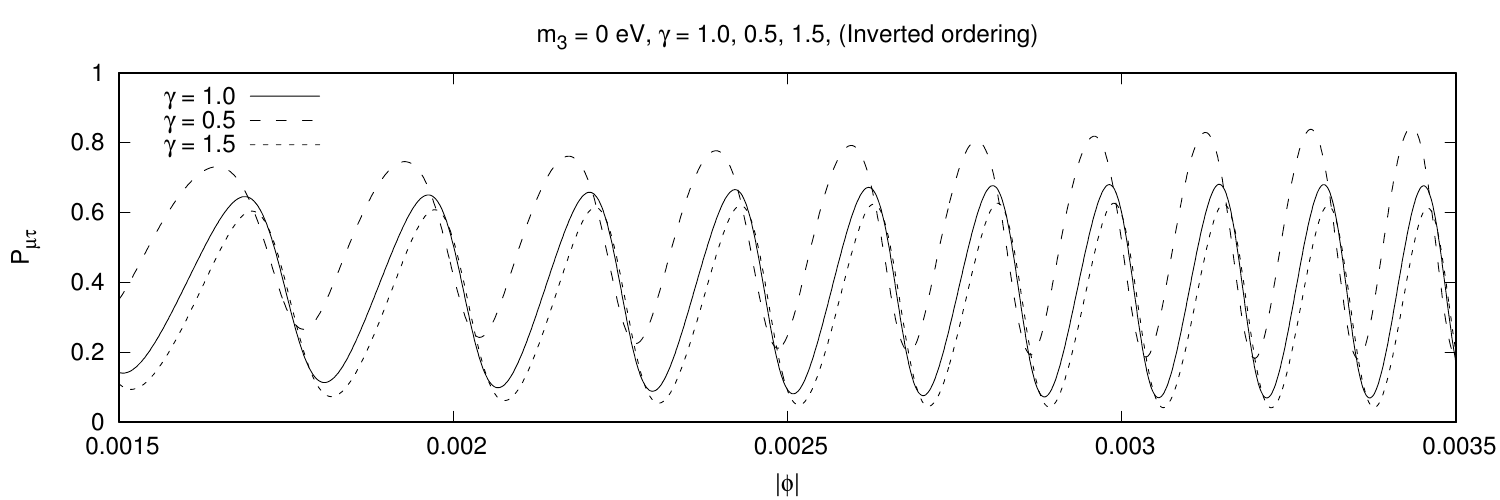}
	\end{center}
    \vspace{-0.5cm}
	\caption{Oscillation probability for the three flavor case in inverted hierarchy  with $ m_3 = 0 $. Top panel: Probability of $ \nu_e \rightarrow \nu_\mu $ conversion for $ \gamma = 1.0 $ (solid line), $ \gamma = 0.5 $ (dashed line), $ \gamma = 1.5 $ (dotted line). Mid panel: Probability of $ \nu_e \rightarrow \nu_\tau $ conversion for $ \gamma = 1.0 $ (solid line), $ \gamma = 0.5 $ (dashed line), $ \gamma = 1.5 $ (dotted line). Bottom panel: Probability of $ \nu_\mu \rightarrow \nu_\tau $ conversion for $ \gamma = 1.0 $ (solid line), $ \gamma = 0.5 $ (dashed line), $ \gamma = 1.5 $ (dotted line). We take $ r_D = 10^8 $ km, $ r_S = 10^5r_D $, $ M_{\rm ADM} = 1 M_\odot $ and $ E_0 = 10 $ MeV. Neutrino mass squared differences, mixing angles and the Dirac CP phase are taken from the latest global fit \cite{Esteban:2020cvm}. \label{fig9} }
\end{figure*}

\begin{figure*}[t]
	\begin{center}
		\includegraphics[width=17cm]{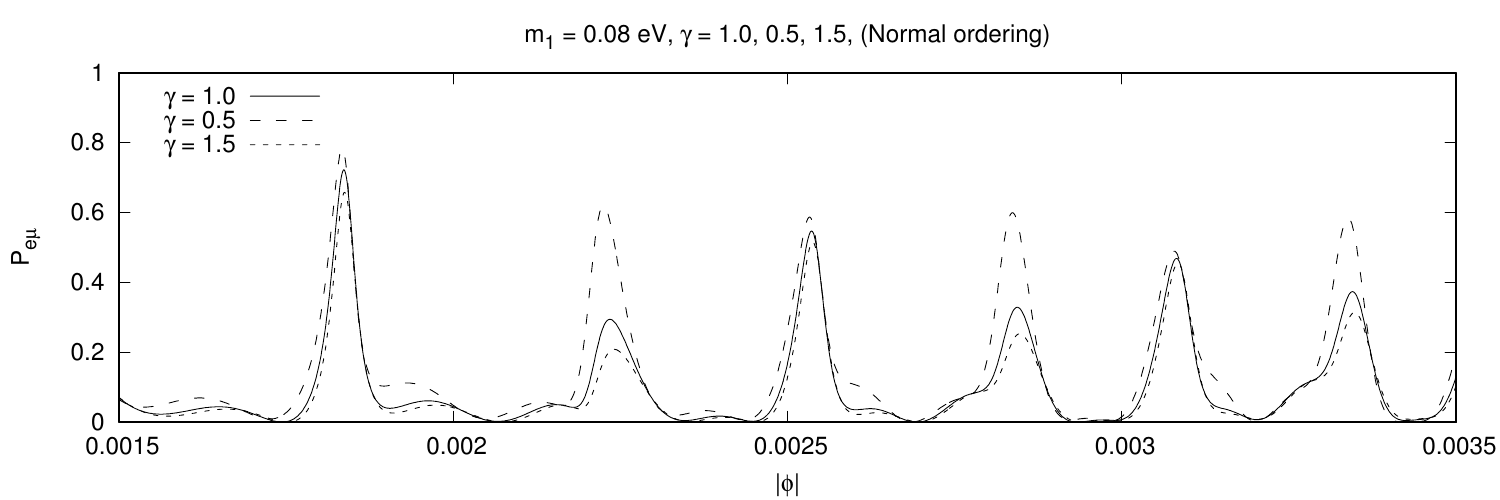}
		\includegraphics[width=17cm]{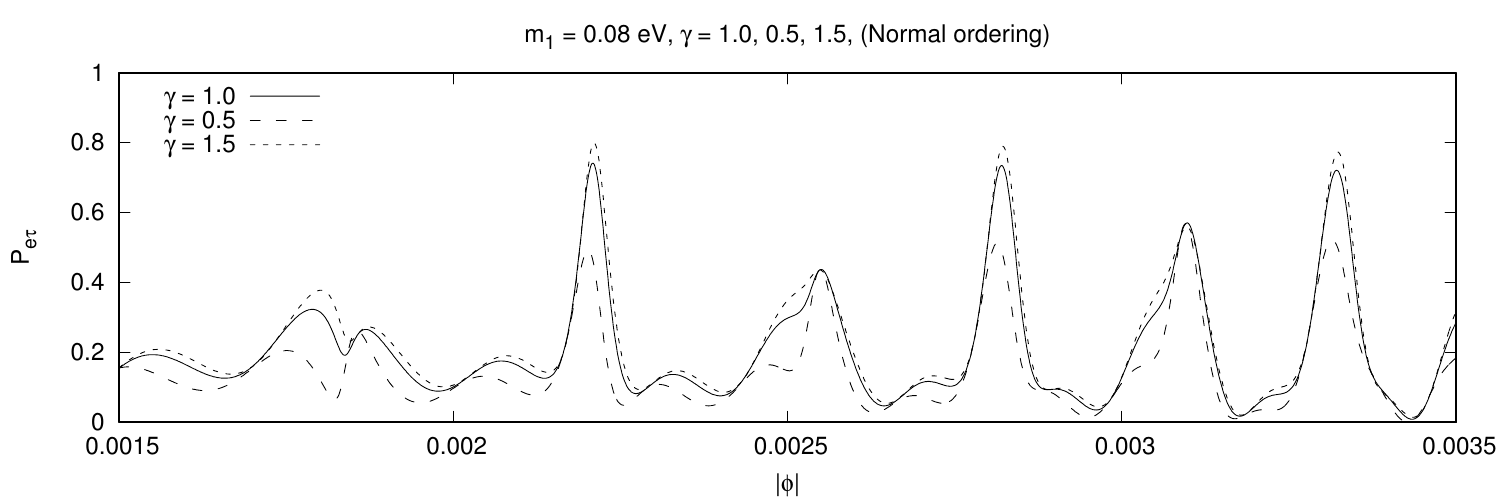}
		\includegraphics[width=17cm]{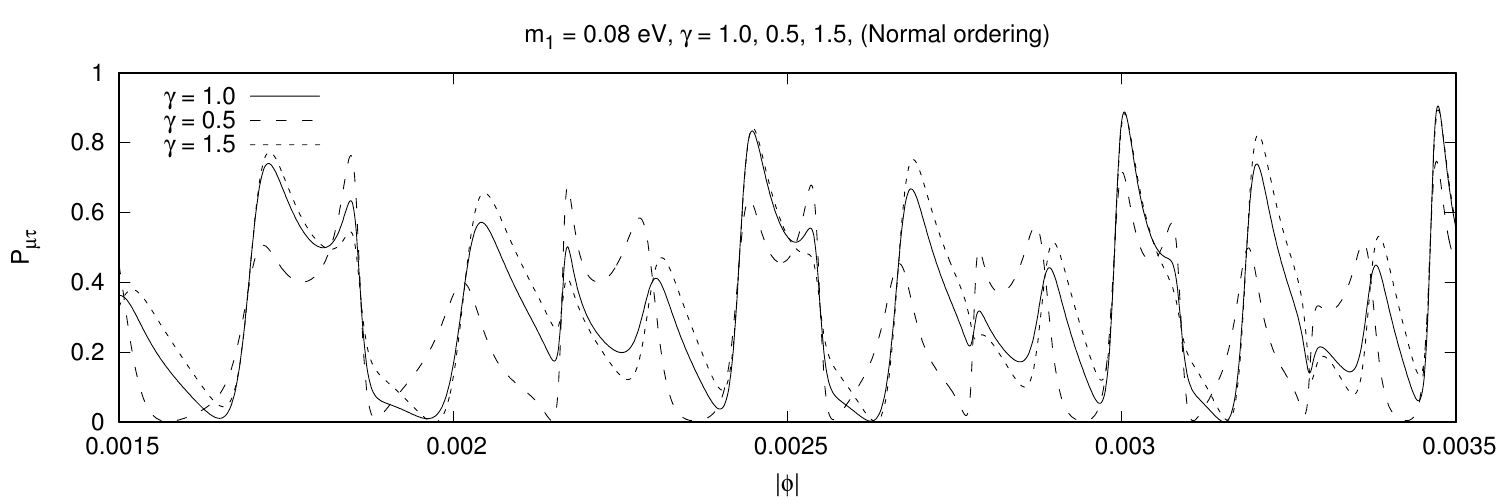}
	\end{center}
	\vspace{-0.5cm}
	\caption{Oscillation probability for the three flavor case in normal hierarchy with $ m_1 = 0.08 $ eV. Top panel: Probability of $ \nu_e \rightarrow \nu_\mu $ conversion for $ \gamma = 1.0 $ (solid line), $ \gamma = 0.5 $ (dashed line), $ \gamma = 1.5 $ (dotted line). Mid panel: Probability of $ \nu_e \rightarrow \nu_\tau $ conversion for $ \gamma = 1.0 $ (solid line), $ \gamma = 0.5 $ (dashed line), $ \gamma = 1.5 $ (dotted line). Bottom panel: Probability of $ \nu_\mu \rightarrow \nu_\tau $ conversion for $ \gamma = 1.0 $ (solid line), $ \gamma = 0.5 $ (dashed line), $ \gamma = 1.5 $ (dotted line). We take $ r_D = 10^8 $ km, $ r_S = 10^5r_D $, $ M_{\rm ADM} = 1 M_\odot $ and $ E_0 = 10 $ MeV. Neutrino mass squared differences, mixing angles and the Dirac CP phase are taken from the latest (NuFIT 5.0 (2020)) global fit \cite{Esteban:2020cvm}. \label{fig10} }
\end{figure*}

\subsection{Three flavor case: numerical results}\label{sec:3flav}

In this section, we shall apply the formalism developed earlier to the realistic three flavor neutrino model. We start with Eq.~\eqref{e-prob-gen} and write down the oscillation probability expressions for $ \nu_e \rightarrow \nu_\mu $, $ \nu_e \rightarrow \nu_\tau $ and $ \nu_\mu \rightarrow \nu_\tau $ conversion. The $ 3\times3 $ matrix $ U $ is the usual PMNS matrix parametrized by three angles $ \theta_{12} $, $ \theta_{13} $, $ \theta_{23} $ and the Dirac CP phase $ \delta_{\rm CP} $. Similar to the two flavor case, we solve the quartic polynomial in Eq.~\eqref{eq-quart} numerically to find the lensing locations and calculate the different oscillation probabilities in these locations. The numerical values of the parameters here are taken from the latest global best fit of neutrino oscillation data \cite{Esteban:2020cvm}. The parameters are $ \theta_{12} = 33.44^\circ $ ($ \theta_{12} = 33.45^\circ $), $ \theta_{13} = 8.57^\circ $ ($ \theta_{13} = 8.61^\circ $), $ \theta_{23} = 49.0^\circ $ ($ \theta_{23} = 49.3^\circ $), $ \Delta m_{21}^2 = 7.42 \times 10^{-5} \; {\rm eV}^2 $, $ \Delta m_{31}^2 = 2.514 \times 10^{-3}\; {\rm eV}^2 $ ($ \Delta m_{32}^2 = -2.497 \times 10^{-3}\; {\rm eV}^2 $) and $ \delta_{\rm CP} = 222^\circ $ ($ \delta_{\rm CP} = 286^\circ $) for normal (inverted) hierarchy.         

The results for conversion probability in three flavor neutrino oscillation are shown in Figures \ref{fig7}-\ref{fig10}. $ \mathcal{P}_{e\mu} $, $ \mathcal{P}_{e\tau} $ and $ \mathcal{P}_{\mu\tau} $ for $ \gamma = 1 $ and $ m_1 = 0 \; {\rm eV} $ is plotted in Figure \ref{fig7}. Figure \ref{fig8} shows the $ \gamma $ dependence of the conversion probabilities when the lightest neutrino is massless $ m_1 = 0 \; {\rm eV} $ for normal hierarchy. The top, mid and the bottom panel shows $ \mathcal{P}_{e\mu} $, $ \mathcal{P}_{e\tau} $ and $ \mathcal{P}_{\mu\tau} $, respectively, for $ \gamma = 1, 0.5 $ and $ 1.5 $. Figure \ref{fig9} shows the $ \gamma $ dependence of the conversion probabilities when the lightest neutrino is massless $ m_3 = 0 \; {\rm eV} $ for inverted hierarchy. The top, mid and the bottom panels show $ \mathcal{P}_{e\mu} $, $ \mathcal{P}_{e\tau} $ and $ \mathcal{P}_{\mu\tau} $, respectively, for $ \gamma = 1, 0.5 $ and $ 1.5 $. Significant differences can be noticed between the normal and inverted hierarchies from comparing Figures \ref{fig8} and \ref{fig9}. Figure \ref{fig10} shows the $ \gamma $ dependence of the conversion probabilities when the lightest neutrino is massive $ m_1 = 0.08 \; {\rm eV} $. Similarly to the two flavor case, we notice that the presence of the parameter $ \gamma $ induces significant effects in three flavor case as well. This is especially noticeable when the lightest neutrino is massive, when we observe the most significant difference in the conversion probabilities for different values of $ \gamma $.   


\section{\label{sec:conc}Conclusion}


Neutrino flavor oscillation is an interesting phenomenon which has the potential to resolve several longstanding questions in fundamental physics. In flat spacetime, with or without matter effects, the flavor oscillation depends only on the difference of the squared masses and does not give us any information about individual neutrino masses. However, neutrino flavor oscillation in a curved spacetime depends on the absolute neutrino masses, as shown in several earlier works, and may be relevant for neutrinos produced by highly energetic astrophysical phenomena. In particular, the effects of gravitational lensing on neutrino oscillations appear to be promising in determining the neutrino masses \cite{Swami:2020qdi}.

This dependency on the individual neutrino masses exists for a wide class of trajectories, including of course geodesics, thus providing a novel potential experimental method to determine masses through the measurement of flavor transition probability. In this work, we extended this method to a larger class of geometries which include mass quadrupole moment by considering a static, axially symmetric and asymptotically flat solution of Einstein's field equations in vacuum known as $ \gamma $-metric. The additional parameter $ \gamma $ in the line element describes deviations of the source from spherical symmetry. We showed how deviations from the usual Schwarzschild metric for the background geometry in which neutrinos travel affects the flavor transition probability and thus potential observations.   
Conversely, if one knows the transition probabilities, this method provides in principle a new and unique tool to measure $ \gamma $ and hence the quadrupole moment of a spacetime which would otherwise be impossible by usual weak optical lensing experiments.  

The numerical results of oscillation probability in the $\gamma$-spacetime for two and three flavor neutrinos
show that the probability increases for decreasing $ \gamma $ at certain locations of the detector on the orbit. This is true for both normal and inverted hierarchies when the lightest neutrino is massless, 
and changes in the value of $\gamma$ alter the peak of the oscillation probability. 
Similar qualitative but different quantitative behaviour is observed when the lightest neutrino is massive showing that the effects due to the absolute neutrino mass depend also on the value of $ \gamma $.
This suggest the possibility of measuring the neutrino masses if the quadruple moment of the source is known or, vice versa, measuring the quadrupole moment if the oscillation probabilities are known.


While the toy model presented here can be considered as a first step, 
a more realistic approach that may be pursued in the future will require the use of a quantum field theoretical treatment \cite{Grimus:2019hlq, Capolupo:2020wlx} and the choice of a geometry that describes the exterior of rotating compact objects. Additionally, to have a complete picture of gravitational effects on neutrino oscillations, one would need to consider effects of gravitational decoherence \cite{Swami:2021wbf}, gravitational spin-flip transitions of neutrinos \cite{Sorge:2007zza, Dvornikov:2019sfo} etc. We would also like to point out some earlier work \cite{Blasone:1995zc,Blasone:1998hf} where it was shown using a QFT-based formula that neutrino oscillations (in vacuum-flat spacetime) in the ``non-relativistic" regime depends on individual neutrino masses.

In the future the growth of observations of neutrinos from distant highly energetic astrophysical sources will play an important role in our understanding of compact objects in the universe as well as fundamental physics. In particular, gravitational lensing of neutrinos can provide a unique probe to test the nature of the source and the validity of modified gravity theories. 
In this context, detecting neutrinos from extra-solar sources lensed by compact objects is expected to provide constraints on the geometry ond thus the nature of the lensing object. 
In future work we plan to extend the framework to modified theories of gravity to determine how future neutrino observations may be used to test alternative theories. 
The hope is that the growth of multi-messenger astronomy \cite{Meszaros:2019xej} and extra-galactic neutrino astronomy \cite{IceCube:2018cha}, will make it possible to test the nature of the geometry surrounding compact objects, put constraints on the validity of modified theories of gravity and explore fundamental properties of neutrino physics such as their absolute masses, which remain unknown with present oscillation experiments.


\noindent
\acknowledgements
The authors thank Dr. Yong Tang for careful reading of the manuscript and for useful comments. HC acknowledges support from University of Chinese Academy of Sciences through Special Research Assistant fellowship (SRA). DB acknowledges the support from Early Career Research Award from DST-SERB, Government of India (reference number: ECR/2017/001873). AA and BA acknowledge  Grants of the Uzbekistan Ministry for Innovative Development and Grant NT-01 of the Abdus
Salam International Centre of Theoretical Physics. AA also acknowledges the support from Chinese academy of sciences through PIFI foundation.










\end{document}